\documentclass[singlecolumn]{emulateapj}
\usepackage{float,epsfig}
\usepackage{natbib}
\newcommand{\LA}{\mbox{\raisebox{-0.6ex}{$\stackrel{\textstyle<}{\sim}$}}}
\newcommand{\cxo}{{\sl Chandra}}
\newcommand{\xmm}{{\sl XMM-Newton}}
\newcommand{\ngc}{{NGC~2903}}
\newcommand{\msun}{$M_{\odot}$}
\newcommand{\ergl}{erg~s$^{-1}$}

\newcommand{\hi}{H{\sc i}}

\newcommand{\ha}{H$\alpha$}
\newcommand{\hii}{H{\sc ii}}
\newcommand{\nh}{$N_{\rm H}$}
\newcommand{\nhkt}{$N_{\rm H}-kT_e$}

\newcommand{\sst}{{\sl Spitzer}}
\newcommand{\ros}{{\sl ROSAT}}
\newcommand{\galex}{{\sl GALEX}}

\newcommand{\um}{$\mu$m}

\newcommand{\msfr}{M$_{\odot}$~yr$^{-1}$}

\slugcomment{Accepted to Astrophysical Journal}

\begin{document}

\title{Hot Diffuse Emission in the Nuclear Starburst Region  of NGC~2903}

\author{Mihoko~Yukita\altaffilmark{1}, Douglas~A.~Swartz\altaffilmark{2},
Allyn~F.~Tennant\altaffilmark{3}, Roberto~Soria\altaffilmark{4}, and
Jimmy~A.~Irwin\altaffilmark{1}}

\affil{\altaffilmark{1} Department of Physics and Astronomy, University of Alabama, Tuscaloosa, AL 35487, USA \\
\altaffilmark{2} Universities Space Research Association,
    NASA Marshall Space Flight Center, ZP12, Huntsville, AL 35812, USA\\
\altaffilmark{3}  Space Science Office,
    NASA Marshall Space Flight Center, ZP12, Huntsville, AL 35812, USA\\
\altaffilmark{4}    International Centre for Radio Astronomy Research, Curtin University, GPO Box U1987, Perth, WA 6845, Australia}

\begin{abstract}
We present a deep \cxo\ observation
of the central regions of the late-type barred spiral galaxy NGC 2903.  
The \cxo\ data reveal  soft ($kT_e$ $\sim$ 0.2--0.5~keV) diffuse emission in the nuclear starburst region and extending
 ~$\sim$2\arcmin\ ($\sim$5~kpc)  to the north and west of the nucleus.
Much of this soft hot gas is likely to be  from local active star-forming regions; however, besides the nuclear region,
 the morphology of hot gas does not strongly correlate with the bar or other known
 sites of  active star formation. 
 The central $\sim$650~pc radius starburst zone
exhibits much higher surface brightness diffuse emission  than the surrounding regions
 and a harder spectral component in addition to a soft component
 similar to the surrounding zones.
 We interpret the hard component as being also of thermal origin with $kT_e\sim$3.6~keV and to be
 directly associated with a wind fluid produced by supernovae and massive star winds  
 similar to the hard diffuse emission seen in the starburst galaxy M82. 
The inferred terminal velocity for this hard component,
  $\sim$1100~km~s$^{-1}$, exceeds the local galaxy escape velocity
  suggesting a potential outflow into the halo and possibly escape from the galaxy gravitational potential.
Morphologically, the softer extended emission from nearby regions does not
 display an obvious outflow geometry.
However, the column density through which the X-rays are transmitted is 
 lower in the zone to the west of the nucleus compared to that from the east
 and the surface brightness is relatively higher suggesting some of the soft hot gas
 originates from above the disk; viewed directly from the western zone but through
 the intervening disk of the host galaxy along sightlines from the eastern zone.
There are several point-like sources
 embedded in the strong diffuse nuclear emission zone.
Their X-ray spectra show them to likely be compact binaries.
None of these detected point sources 
 are  coincident with the mass center of the galaxy
 and we place an upper limit luminosity from any point-like nuclear source
 to be $<$2$\times$10$^{38}$~\ergl\ in the 0.5$-$8.0~keV band which indicates that 
NGC~2903 lacks an active galactic nucleus.  Heating from the nuclear starburst and a galactic wind 
may be responsible for preventing cold gas from accreting onto the galactic center.
\end{abstract}


\section{Introduction}

Stellar bars are common features of disk galaxies \citep{masters11,sellwood93, 
buta11}
where they efficiently channel gas into the central regions.
The dynamics of bars is critical to galactic (pseudo)bulge growth
and hence to evolution along the Hubble sequence \citep{kormendy04}
and to fueling central supermassive black holes \citep[e.g.,][]{shlosman90}
in isolated galaxies.
In isolated late-type disk galaxies,
stellar bars can efficiently transport gas into their central regions because 
of
their relatively low bulge masses compared to early-type disks.
Gas may accumulate at the ends of the bar before being driven inward by the
gravitational torques within the bar \citep{combes93}.
The gas forms thin ``centered'' dust lanes along the bar
\citep{athanassoula92}.
There is less shear in the bars of late-type disks
giving rise to substantial star formation along the bar.

There are often no central resonances
in late-type disks, in contrast to bars in early-type disks;
the bar extends to about the turn-over radius of the rotation curve, near
the inner Lindblad resonance, and corotation occurs even further out in the 
disk \citep[e.g.,][]{shlosman99}.
In this case, accretion does not stall to form a circumnuclear ring
but accumulates until self-gravity leads to burst-like episodes 
\citep{sarzi07} of star
formation~--~inflow may be impeded by stellar feedback but not by
resonance phenomena as in early-type disks.
Star formation and feedback can deplete the gas reservoir even at small
distances beyond the sphere of influence of central black holes
\citep{davies10,thompson05}
or, it may enhance fueling of black holes if the starburst occurs very
near the central compact object \citep{watabe08}.

The study of structures such as bars and circumnuclear rings in X-ray light
 can directly reveal details that are hidden at many other wavelengths.
The X-ray morphology not only helps to pinpoint locations of recent star
 formation but also traces high-pressure
 hot gas wind fluid that may quench star
 formation, impede accretion toward the galactic center, and drive
 energetic outflows.
The best subjects for X-ray studies of these phenomena are isolated
 nearby galaxies inclined enough to distinguish galactic outflows
 while leaving the nucleus exposed to direct view and galaxies lacking
 strong active galactic nucleus (AGN) activity which could mask the fainter X-ray emission from
 circumnuclear star formation.

\begin{figure*}
\begin{center}
\includegraphics[width=0.7\textwidth]{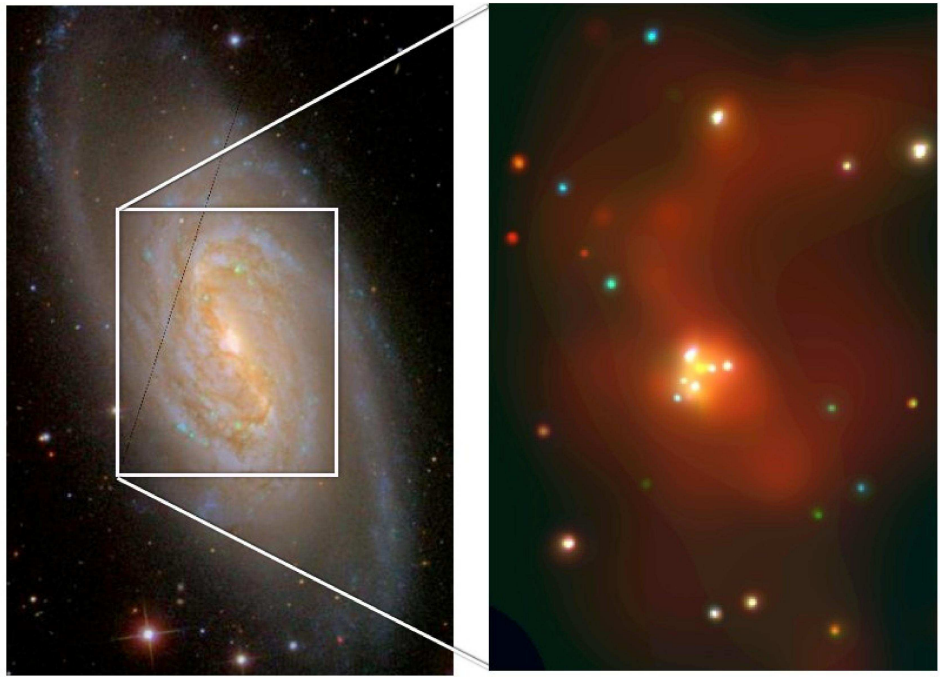} 
\caption{Left: the central 4\arcmin$\times$ 6\arcmin\ 
region of SDSS image.  Blue, green, and red correspond to the
$g$, $r$, and $i$ bands, respectively. Young stars in spiral arms are traced in blue, and bright \hii\ regions appear
in green. The complex dust lane structures  along the bar and spiral arms are also noticeable
in darker color. White box corresponds to the region shown in the right panel.  Right: the 
central 2\arcmin$\times$3\arcmin~region of \cxo\ image. Red, green, and blue correspond to
the energy ranges, 0.5--1.0~keV, 1.0--2.0~keV, and 2.0--8.0~keV, respectively.  The image is adaptively smoothed with
a minimum signal-to-noise of 3.
  \label{f:sdssx}} 
\end{center}
\end{figure*}

NGC~2903 is a nearby \citep[8.9~Mpc, 1\arcsec$=$43~pc;][]{drozdovsky00}
late-type barred SAB(rs)bc galaxy with strong circumnuclear star formation.
NGC~2903 was first classified as having a central concentration of
multiple bright hotspots superposed on a fainter background
in an early galaxy classification scheme \citep{morgan58} based on central 
light morphologies
\citep[see also][]{sersic65,sersic67}.
These hotspots have since been identified as young massive star clusters
in optical and near-IR images interspersed with numerous bright \hii\ regions
\citep{planesas97,alonso01,hagele09}
within an $\sim$650~pc diameter circumnuclear star-forming region.
The star formation rate (SFR) in this region alone is $\sim$0.7~\msfr\ 
\citep{alonso01,leon08}.
This nuclear region is bright at wavelengths from X-rays
\citep{tschoke03,perez10}, UV \citep[e.g.,][]{popping10}, optical and
mid-IR, to CO \citep{sheth02,helfer03}, HCN \citep{leon08}, and radio 
\citep{wynn85,tsai06}.
For the most part, the nuclear region displays a patchy morphology
with little obvious symmetry.
Perhaps the precise center is best seen in high-resolution $H$-band images
\citep[e.g.,][]{perez00,alonso01}
 where a
two-arm nuclear spiral pattern is also evident.

The nuclear region of NGC~2903 is clearly being fueled by inflow of gas along 
its bar
\citep{hernandez05,leon08}.
The total SFR along the bar is comparable to the rate in the
nuclear region, $\sim$0.7~\msfr\ \citep{leon08,popping10}.
The stellar bar extends roughly 1\arcmin\ ($\sim$2.5~kpc)
from the nucleus at a P.A. of
24\arcdeg\
\citep{sheth02}.
A strong, nearly unbroken dust lane along the bar is clearly visible in 
molecular bands
\citep{sheth02,leon08}.
The dust lane leads the stellar bar (P.A. $\sim$30\arcdeg,
rotation is counterclockwise (CCW) and the disk is inclined $\sim$60\arcdeg)
but trails active star formation visible as a curved string of
numerous bright \hii\ regions further downstream
\citep[][see particularly Figure~1 of Sheth et~al.]{sheth02,popping10,knapen02}.
These offsets are expected inside the corotation radius
when gas and dust dissipates energy and loses angular momentum as it
passes through the stellar bar \citep[e.g.,][]{martin97}.

The bar is not as distinct at shorter wavelengths but instead appears patchy 
as does the
outer flocculent spiral arm structure \citep{popping10}.
The spiral pattern is clearest in \hi\ maps \citep[e.g.,][]{walter08}.
Deep images reveal an \hi\ envelope extending to at least three times the 
optical
diameter of the galaxy \citep{irwin09}
indicating no recent interactions which tend to truncate the \hi\ disk
\citep{higdon98,chung09}. 

Hence, NGC~2903 is in many respects a textbook
example of an isolated late-type barred spiral galaxy
displaying a gas-rich star-forming nucleus and bar structure.
This makes the galaxy an ideal target to study gas flow driven by the
 internal structures. 
Here we investigate the properties of the X-ray emitting hot gas and
point-like
sources in NGC~2903 revealed by a deep \cxo\ {\it X-ray Observatory} 
spectrophotometric image.
We mapped X-ray emission in order to examine 
the correlation of the hot gas and point sources with sites of 
current star formation
and of the various galactic structures in \S~\ref{sec:xmap}.
The detailed properties of the hot gas is analyzed in \S~\ref{sec:diffuse}, where
we do confirm the emission asymmetry in the diffuse X-ray light first
revealed by {\it ROSAT}/PSPC images \citep{tschoke03} and visible in 
contemporary
\xmm\ data \citep{perez10}.
We search for an nuclear activity in \S~\ref{sec:nsb} and find that
none of the point sources are coincident with the nucleus of the galaxy.
We discount the possibility of a low-luminosity AGN in NGC~2903
\citep[cf.][]{perez10}.
We discuss that some of this emission likely traces a galactic wind or fountain
extending above the disk of NGC~2903 originating in the nuclear hotspot 
region and revealing active stellar feedback and its influence on the growth of
nucleus in \S~\ref{sec:dis}.

\section{Mapping the X-ray Emission from NGC~2903}\label{sec:xmap}

NGC~2903 was observed with \cxo\  using the ACIS-S instrument in imaging mode on
 7~March 2010  (ObsID 11260).
The level 1 event list was reprocessed using the {\footnotesize{CIAO}} (version 4.3) tool {\it
  acis\_process\_events} to apply the CTI correction and to
adjust for the time-dependent gain (with {\footnotesize{CALDB} } 4.3.0).
Then, bad and hot pixels as well as bad status bits and grades were
 filtered out to create a level 2 event file.
In order to identify any background flares, a lightcurve, binned to 500~s increments,
 was created from events in the energy range of 10$-$12~keV where the telescope
 mirrors were insensitive to X-rays.
Background flares were identified and removed using  the 3$\sigma$ clipping 
method\footnote{http://cxc.harvard.edu/ciao/threads/flare/}
  which resulted in  92~ks of the final GTI for
 this observation.
The angular resolution of the \cxo\ mirrors is slightly better
 than the pixel resolution of the ACIS detector.
Hence, we created a subpixel resolution image in the nuclear region  by  applying the  sub-pixel event-repositioning algorithm ``EDSER'' of
 \citet{li04}, which became available in {\footnotesize CIAO} v.4.3.
For the rest of the galaxy, the standard 0\farcs49 pixel resolution image
was created.

X-ray emission in star-forming galaxies like \ngc\ 
 arises from various sources
   including bright X-ray binaries (appearing as 
   relatively hard point-like sources),
   diffuse hot gas from supernovae and
   massive star winds, and underlying emission from numerous fainter
   unresolved X-ray binaries.
Figure~\ref{f:sdssx} displays a 4\arcmin$\times$6\arcmin\
 3-band optical image of \ngc\
 obtained from the Sloan Digital Sky Survey \citep{york00}
 along with a 3-color smoothed X-ray image of the
 central 2\arcmin$\times$3\arcmin\ region.
The X-ray image shows numerous point-like sources as well as
 high surface brightness relatively hard diffuse emission
 in the nuclear region and more extended and softer diffuse emission
 throughout much of this central region of \ngc.
In this section, we first characterize  the bright detected point sources and then describe the
 morphology of the diffuse emission and create maps of the thermodynamic
 state of the hot gas from which we identify larger zones  that we analyze  spectroscopically
in \S~\ref{sec:diffuse}.

\subsection{X-Ray Point Sources}\label{S:pts}

The source finding tool in {\it lextrct} \citep{tennant06} was applied
 in the energy range of 0.5$-$8.0~keV to detect  point sources inside the
 $D_{25}$ isophote.
A total of 92 point-like sources were detected with a
 signal-to-noise ratio (S/N) above 2.4 (see Tennant 2006) and with a 
  minimum of 5 counts above the background uncertainty.
These sources are tabulated in the Appendix where we
 also provide spectral analysis of the brightest sources.
A spectrum obtained by co-adding events from all sources
 was also examined. Its rough spectral shape is that of an
  absorbed power law with spectral index $\Gamma=1.95 \pm 0.02$
 although there are some large fit residuals in the $\sim$1$-$2~keV 
 range likely due to contributions from (point-like) thermal emission sources.
The total luminosity in point sources is 1.9$\times$10$^{40}$~\ergl\
 in the 0.5$-$8.0~keV range.
 
Visual inspection of the point source locations overlaid
 on the \galex\ UV, SDSS optical, \sst\ near-IR, and
 the BIMA CO images revealed very few correlations with
 features visible at these other wavelengths.
The notable exceptions are (1) the nuclear region which contains
  8 point-like X-ray sources within a 15\arcsec\ (650~pc) radius
(this region was unresolved in previous \xmm\ and \ros\ X-ray observations)
 and (2) a bright object north of the nucleus,
 designated CXOUJ093209.7+213106,
 coincident with strong emission from the UV to mid-IR bands.

We note that there is also strong, relatively hard, diffuse emission
 extending throughout the nuclear region.
Some detected point sources in the nuclear region may be knots of
 hot gas rather than X-ray binaries traditionally associated with
 point-like sources.
We inspected the X-ray colors and sizes of these sources to help
 differentiate compact binaries from concentrations of hot gas
 following the simple methods in \citet{swartz06}.
For the latter, we applied two-dimensional Gaussian models to the distribution 
 of X-ray events to determine if sources are point-like
 indicating compact binaries or more extended hot gas concentrations.
None of the 8 point-like sources in the nuclear region were conclusively
 identified as having a thermal origin.
Further analysis of the point sources and diffuse emission in the central
 15\arcsec\ region
 is deferred to \S~\ref{sec:nsb}.

\subsection{Diffuse X-Ray Emission}\label{sec:map}

We confine our analysis of the diffuse X-ray emission in \ngc\ to the
 2\arcmin$\times$3\arcmin~region shown in the right panel of Figure~\ref{f:sdssx}.
 There is no evidence of diffuse emission beyond this region
in either the \cxo\ or the recent \xmm\ observations.
A point-source-free diffuse emission map was created by removing the
detected point sources and then filling the excluded regions by
 sampling the Poisson distribution of the mean levels of their local
 backgrounds using the {\footnotesize{CIAO}} tool {\it dmfilth}.  

Figure~\ref{f:xdiff} shows this
 diffuse emission map
 in the soft (0.5$-$2.0~keV) X-ray band.
This emission map has been smoothed with the {\footnotesize {CIAO}} tool {\it aconvolve}
  using a Gaussian smoothing function with a width of 10~pixels ($\sim$5\arcsec).
Also shown for comparison is an unsmoothed image binned by 4 pixels
($\sim$2\arcsec) which is the 
pixel scale used below for building maps of the thermodynamic state of the gas.
Diffuse emission is clearly present in the central region.
The low surface brightness diffuse emission extends to a few minutes of arc, 
 especially to the north  of the nucleus.

\subsubsection{Galactic Structures and X-Ray Morphology}\label{sec:multi}

\begin{figure*}
\begin{center}
\includegraphics[width=0.7\textwidth]{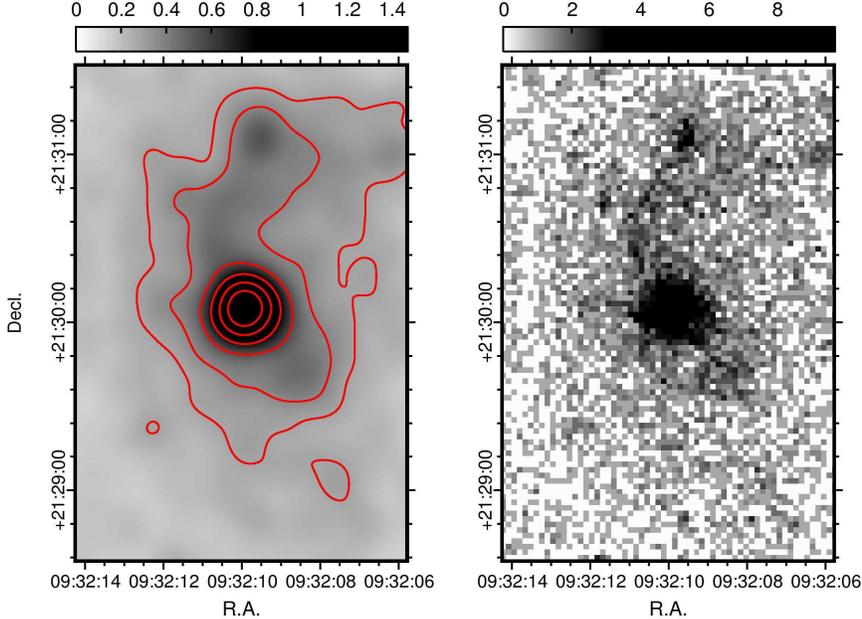} 
\caption{Left: the central 2\arcmin
  $\times$3\arcmin\ of the point source-free X-ray soft
  (0.5$-$2.0~keV) diffuse emission map.  The image was
  smoothed using the {\it CIAO} tool, {\it aconvolve} with 
  Gaussian $\sigma$ width of 5\arcsec. 
  The contours are also plotted in square root scale,
 specifically 0.075, 0.132, 0.307,0.59, 1.0 and 1.52 count~pixel.
 The pixel scale of this image is 0\farcs492.
 Right: the same as the left image, besides binned by 2\arcsec\ scale
rather than smoothed.  The color bars for the both images indicate
 the square root of the intensity with the unit of count.
\label{f:xdiff}} 
\end{center}
\end{figure*}

\begin{figure*}
\begin{center}
\includegraphics[width=0.8\textwidth]{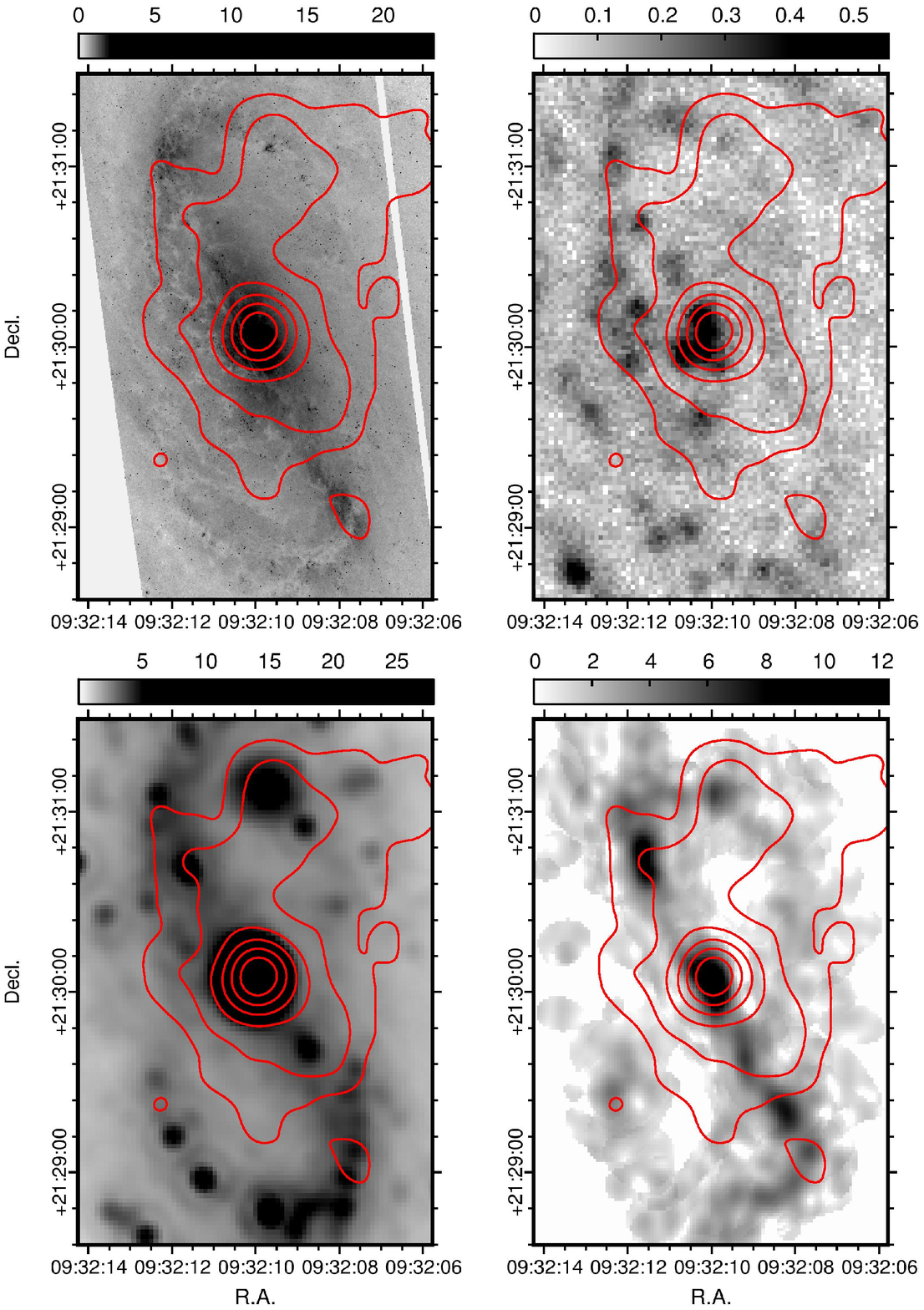}
\caption{Central 2\arcmin $\times$3\arcmin\ of the NGC~2903 of
 {\it HST}/ACS F814W (top left), \galex\ FUV (top right), CO (bottom right), and \sst\
 24~\um\  (bottom right) images. The color bars indicate a square root of intensities.
 The {\it HST}  image has the pixel scale of 0\farcs05 and the
 intensity  unit of electron; 
 1$e$ corresponds to a fluence of
 7.1$\times$10$^{-20}$~erg~cm$^{-2}$~\AA$^{-1}$, that is a specific flux 
 $F_{\lambda}$= 1.0$\times$10$^{-22}$~erg~s$^{-1}$~cm$^{-2}$~\AA$^{-1}$.
The pixel scale of the \galex\ FUV image is 1\farcs5 with the
intensity units of
counts~s$^{-1}$. The BIMA SONG CO image has the pixel scale of 1\arcsec.
The unit of image is Jy~(beam$)^{-1}$~km~s$^{-1}$.  
The \sst\ 24~\um\ image has the pixel scale of 1\farcs 5, and the map
intensity is in Jy sr$^{-1}$. 
The contours depict the intensity of the smoothed 
 X-ray diffuse emission shown the left panel of Figure~\ref{f:xdiff}.
\label{f:sf}} 
\end{center}
\end{figure*}

Figure~\ref{f:sf} shows the same 2\arcmin$\times$3\arcmin~region as 
Figure~\ref{f:xdiff} in different wavelength images: {\it HST}/ACS F814W,
 \galex\ FUV, \sst\ 24~\um, and BIMA SONG CO. 
The contours of the smoothed X-ray diffuse emission from Figure~\ref{f:xdiff}
 are overlaid
on each panel of Figure~\ref{f:sf} to compare the diffuse X-ray emission
 to the galactic structures and
star-forming regions visible at these other wavelengths.

The brightest feature at all wavelengths is the nuclear region.
The bar is also clearly represented in the {\it HST}/ACS F814W image
 tracing the stellar emission and in the CO and
24~\um\ images tracing cold molecular gas and dust.
\citet{popping10} and \citet{sheth02} report an offset between these stellar
 and cold gas/dust bars and the \ha\
  emission tracing young high-mass stars
  with the \ha\ emission {\sl leading} the CCW-rotating bars.
The UV image, tracing stars younger than $\sim$100~Myr, appears
 much more spotty and does not follow the bar structure.
\citet{popping10} also note the UV emission is patchy and does not
 correlate well with either the \ha\ or 24~\um\ tracers of current star formation.
Instead, they find extremely young ages ($<$10~Myr) for UV-emitting clusters
 along the bar and much older ($>$100~Myr) for other clusters in the field. 
At the longer wavelengths, the emission extends beyond the ends of the
 bar into the trailing spiral arms and forms a distinct, though patchy,
 S-shaped pattern.

The highest surface brightness X-ray emission also occurs in the central
 region of \ngc.  There is also weaker  emission extending beyond the nucleus in the X-ray image but, except for a 
 few bright knots,
 the emission is not confined to the stellar or molecular bar nor the S-shaped pattern
 visible in the IR and CO bands.
Instead, this low surface brightness emission extends to the northwest
 of the nucleus as if filling the inner portion of the upper half of the S-shaped region of star
 formation traced by the 24~\um\ emission.
Contrary to this,  there is no bar-like morphology along the southern half of the bar
 nor extended emission in the southwest S-shaped region in X-ray.
We considered a difference in the column density might account for this asymmetry; 
 however, no large
 difference in extinction between the north and south halves of the bar is reported
 from \ha\ or 24~\um\ analysis \citep{popping10}.
 In summary,
besides the nuclear region and a few isolated extended knots which may be associated with bright star-forming regions,
 there are no strong correlations between
 the X-ray morphology and the galactic structures seen at various other
 wavelengths.

\subsubsection{X-Ray Thermodynamics Maps}\label{sec:tmaps}

The lack of strong correlations between diffuse X-ray emission surface brightness
 and galactic structures visible at other wavelengths prompted us to
 examine other properties of the diffuse X-ray emission in greater detail.
To this end, we created maps of the thermodynamic state of this 
 hot gas to examine if we see any particular morphological
 features in the X-ray properties.

\begin{figure*}
\begin{center}
\includegraphics[width=0.8\textwidth]{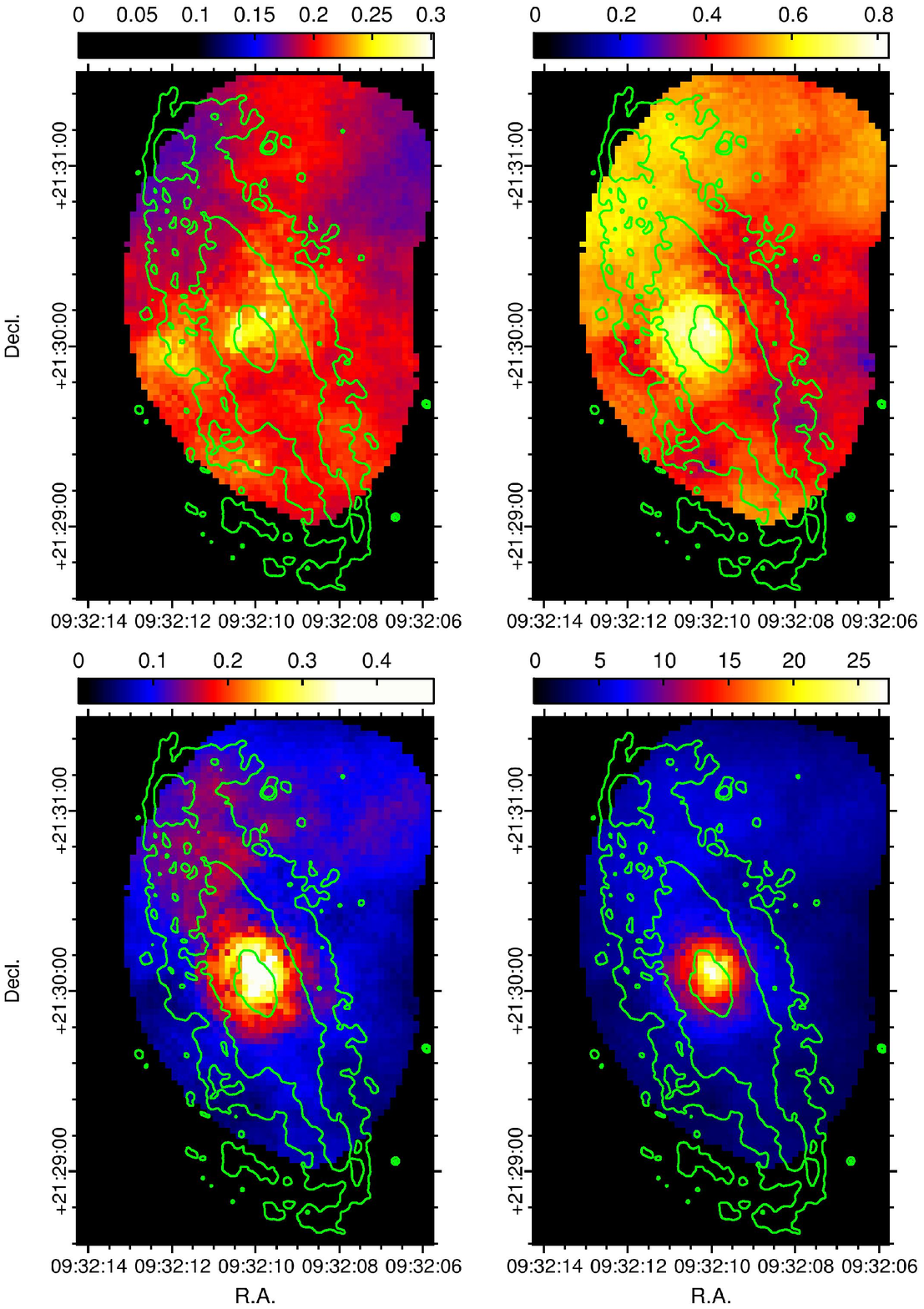} 
\caption{Same region as in Figure~\ref{f:xdiff} of the
temperature (top left), hydrogen column density (top right), electron
 density (bottom left), and pressure (bottom right) of the hot gas derived 
 by applying an absorbed single temperature model (see the text).  The
 units for each map are in keV, 10$^{22}$~cm$^{-2}$, cm$^{-3}$, and 10$^{5}$~K~cm$^{-3}$,
 respectively, as denoted by the color bar at the top of
 each panel.  One pixel corresponds to $\sim$2\arcsec, which is the
 same as the right panel of Figure~\ref{f:xdiff}.  
 However, the spectral extraction regions used to define the pixel
values are larger than 2\arcsec, see the text for details.
 The regions which did not contain enough counts for spectral fitting
 appear black.  We note that a single temperature model
 can result in higher \nh\ and normalization, therefore resulting in a higher density.
  In order to compare with the galactic 
structures, the {\it HST}/ACS F814W contours are overlaid in green.
\label{f:Tmap}} 
\end{center}
\end{figure*}

Specifically, we created a point-source-free image binned by 4 pixels ($\sim$2\arcsec). 
Then, we extracted a spectrum from a circular
 region around each pixel in the binned image. 
This procedure follows
 the method described in \citet{randall08}, who employed the technique
of \citet{osullivan05} and \citet{maughan06}.
The radius of each region was chosen to contain at least 500~counts
  after background subtraction.
A region to the northeast outside of the $D_{25}$ annulus was used for the background.
We constrained the maximum radius of each region to be  $\le$30\arcsec;
 we discard regions needing $>$30\arcsec\ to contain $>$500~net counts.
Overall, the resulting range of radii is 5\arcsec$-$30\arcsec. 
This is much larger than the
 pixel scale representing the resulting parameter values;
 this is in a sense equivalent to ``smoothing'' the parameter maps.

Each spectrum is then fitted using the energy range 
 between 0.5$-$2.0~keV, because there are very few counts
 detected  above 2~keV after the background subtraction.
We fitted with  an absorbed collisionally ionized 
optically thin plasma model, namely  {\tt phabs*} {\tt mekal} in {\footnotesize XSPEC}.\footnote {The {\tt mekal} model was used here because 
the solution is computed ``on-the-fly'' as opposed to
being pre-defined on a finite model parameter grid as does 
the {\tt apec} model so that
it does not suffer from a discretization problem.}
Due to low counts in each spectrum, we use Poisson statistics to fit spectra and
we fixed the metallicity to the solar value
following the \xmm\ Reflection Grating Spectrometer (RGS) analysis of \citet{perez10};
 see also \citet{pilyugin04} and references therein.
Applying the same model to the spectra of each spatial region
 allows us to spatially resolve relative changes in the thermodynamic properties of the gas.

Figure~\ref{f:Tmap} shows the resulting X-ray temperature, $kT_e$,
 absorbing column density, \nh, electron
density, $n_e$, and pressure, $P$, maps of the hot gas over the
same regions as displayed in Figure~\ref{f:xdiff}.
We also overlay the {\it HST}/ACS F814W contours in order to visualize the underlying galactic structures.

The measured temperature shown in Figure~\ref{f:Tmap}
ranges from  0.2 to 0.3~keV with
 errors of 0.03~keV on average.
This is 
 a typical temperature for hot gas (0.1$-$0.7~keV)
 in star-forming regions in normal or starburst
 galaxies \citep[e.g.,][]{tyler04,strickland04}.
The nuclear starburst region has relatively higher $kT_e$ than the surrounding areas.
The highest temperature in the nuclear region (yellow in the figure)
 is elongated in a roughly NW to SE direction nearly perpendicular to the bar
 and extending to the NW. 
This extension to the northwest is also seen in the \xmm\ image presented in
 \citet{perez10}.
 However, there is no large difference in 
 temperature among these regions overall, 
 which is also consistent with what has
 previously been found in the other star-forming galaxies
 \citep[e.g.,][]{swartz06,wang09}. 
 Interestingly,  the temperature
 along the (northern) bar decreases whereas the surface brightness 
 remains roughly constant. 
The temperature is higher to the northwest of the northern bar
 (within the S-shape).

The measured \nh\ is in the range of (2$-$8$)\times$10$^{21}$~cm$^{-2}$,
 which is much larger than the Galactic \nh\ value of
 3$\times$10$^{20}$~cm$^{-2}$ \citep{dickey90} along the line of sight. 
In fact, this is larger than the intrinsic \nh\ value for \ngc\ of
$\sim$5$\times$10$^{20}$~cm$^{-2}$ derived from the H{\sc i} Nearby
Galaxy Survey data
\citep{walter08} estimated using  their Equation~(5).
We note that applying a single temperature model 
to the multi-temperature gas seen in galaxies
often tends
 to result in a higher column density
 because of the need to suppress low-energy emission in the model \citep[e.g.,][]{yukita10}.
Due to the low fitted model temperatures, the \nh\  uncertainties are somewhat large,  about
 $\pm$(0.5$-$1.0)$\times$10$^{21}$~cm$^{-2}$.  
The nuclear region and the dust lane to the east of the nucleus have
the highest \nh.  
The north side of the bar also shows a relatively higher column
 density, and the regions to the west of the nucleus, where fewer and weaker dust lanes are visible in
 the optical {\it Hubble Space Telescope} ({\it HST}) image, have relatively lower \nh\ values.
However, the spectral properties, both \nh\ and $kT$,  along the south side of bar does not show a
 correlation with the underlying galactic structures, similar to the
 lack of correlation with these features in the X-ray surface brightness.

The X-ray electron density, $n_e$, is estimated from the 
 model fit parameters in the following way.
The spectral model normalization, $K$,  is related to the emission integral,
$\int n_e n_{\rm H} dV \sim n_{e}^{2} fV$, where $V$ is the X-ray
emitting volume and $f$ is a volume filling factor.
This volume filling factor represents the fraction of the volume  actually occupied by X-ray emitting gas.
We assume that the number densities of ions, $n_i$, and electrons  are equal.  
This is a reasonable assumption for hydrogen dominated plasma.
 The electron number density, then, can be estimated as $n_e^2 \propto K/(fV)$.

 In order to derive an electron density map, we assume that the hot gas 
 fully occupies ($f\equiv 1$) a cylindrical volume represented by 
 a scale height, $h=200$~pc, of the disk. The electron density then 
 scales linearly with $(fh)^{-1}$.
This is, of course, not the 
 only possible scenario; the gas (or some fraction of the gas) may
 extend above the disk in the form of an outflow. 
This possibility is examined further in \S~\ref{sec:dis}.
 
Using the derived $n_e$ and the flux-weighted mean X-ray temperature $T_e$,
the hot gas pressure $P/k=2n_eT_e$ can be derived.
The pressure is also highest in the nuclear region, approaching 25 $\times$10$^5$~K~cm$^{-3}$,
 and is more centrally-peaked than either the temperature or the electron density alone.
The pressure distribution is smoother beyond the nuclear region 
 approaching a roughly constant value of (5$-$7) $\times$10$^5$~K~cm$^{-3}$.

In summary, Figure~\ref{f:Tmap} shows that
 the highest density, temperature, and pressure region is the nuclear starburst region.
This is also the highest surface brightness region (Figure~\ref{f:xdiff}).
The density is relatively high in the region along and to the northwest of the northern
 portion of the stellar bar.  
We do not see an analogous feature corresponding to the south side of the bar.
Lastly, the pressure map indicates that only the nuclear
starburst region is highly pressurized if we assume an equal scale
 height for all the regions.
In fact, it is possible that the hot gas is flowing out into a
 low density halo region, which would imply a different  morphology of the gas volume than considered here.

\begin{figure*}
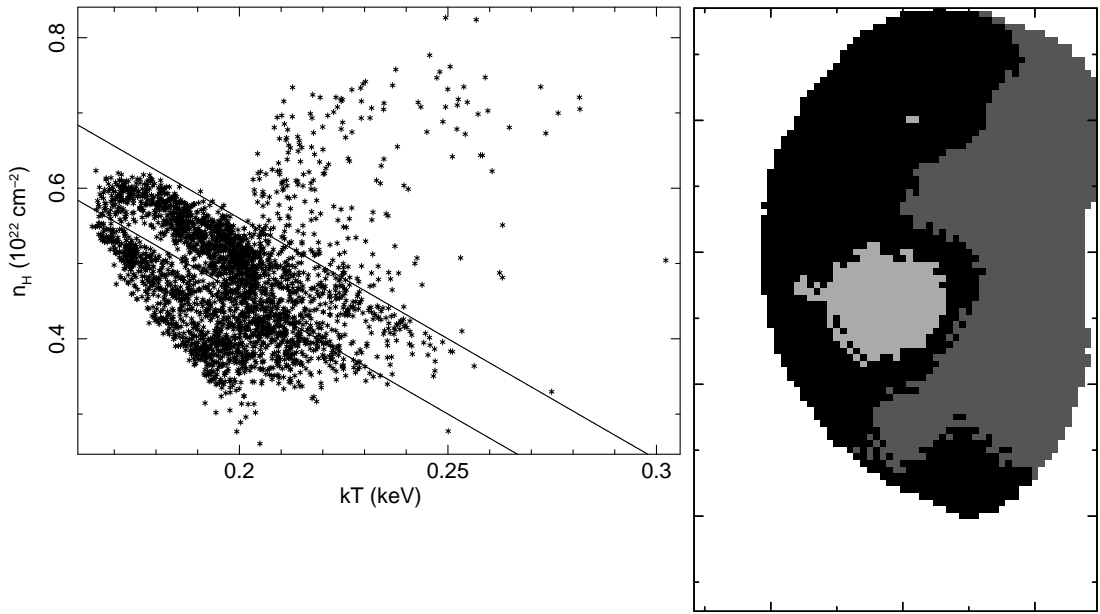

\begin{center}
\includegraphics[angle=-90, width=0.5\textwidth]{f5a.eps} 
\includegraphics[angle=-90, width=0.3\textwidth]{f5b.eps} 
\caption{Left: The \nh\ vs. $kT_e$ plot for the fits of an absorbed
  single temperature model using 0.5$-$2.0~keV.  
In general, there is an anti-correlation between \nh\ and $kT_e$, as a higher
 \nh\ results in a lower $kT_e$ value. We divided the parameter space into
 3 groups as indicated with straight lines. 
Right: the three parameter groups defined in the left
panel are mapped in  spatial
 coordinates. 
The light gray zone indicates the right top group, and
the dark gray zone depicts the bottom left group, and the black zone is the middle
 group in the left panel.
These two plots show that the nuclear region has both high \nh\ and $kT_e$
 values that do not follow the correlation seen in the rest of
 region.
\label{f:banana}} 
\end{center}
\end{figure*}

We point out that the fitted column density and
 temperature parameters are often anti-correlated. 
There are low column density (blue) regions, where the temperature 
is high (yellow and orange regions west of the nucleus).  
Also,  the temperature is lower (red) where the column density is higher (yellow).
Another, though weaker, positive correlation 
sometimes occurs between column density and electron density
(proportional to the square root of the model normalization)
in these single-temperature models.
We show  the column density of all the pixels in the map against their corresponding temperature to
 examine this behavior in Figure~\ref{f:banana}.  

The left panel of Figure~\ref{f:banana} clearly shows that most regions follow
 the anti-correlation with higher \nh\ values tending to correspond to
 lower temperatures.  
These regions are located in the bottom left side of the figure. 
There are also regions that do not follow this trend, 
 spreading into the top right of the plot.
We selected three groups in this parameter space as indicated by lines
 in the left panel of Figure~\ref{f:banana} based on the grouping
 of the regions apparent to the eye.
The right panel of Figure~\ref{f:banana} shows the spatial map of the corresponding
 groups defined in this way.
As anticipated, the nuclear region corresponds to the top right group
 consisting of the hottest, most heavily absorbed gas with clearly  different properties than the other two regions.
The middle and bottom left groups also show strong  spatial correlations. The middle group regions are located to the east and
 the bottom left group are located primarily to the west of the nuclear region.

Although these regions were found using solar abundances, we note
that similar structures are evident when changing abundances to
a different value, such as $Z=0.05~Z_{\odot}$.
In order to understand why these regions
are different,  we analyze the X-ray spectra extracted from these 
 three groupings, or zones, where we are able to apply more complex models
 in the next section.

\section{Further Spectral Analysis of the Diffuse Emission}\label{sec:diffuse}

We have shown in the previous section, using detailed maps of the 
 thermodynamic state of the hot, X-ray-emitting gas, 
 that the spatial complexity observed at other wavelengths 
 is clearly present at X-ray energies.
On the other hand, there are no strong correlations
 between the X-ray morphology and the underlying galactic structure
 with the notable exception of the nucleus.
There is a general trend of lower temperature and lower column density
 to the west of the nuclear region relative to east of the nuclear region.
In this section, we will analyze  in greater detail the spectral properties
of three zones:
the  $\sim$15\arcsec\ nuclear starburst region, and the two extended
 diffuse emission regions to the east and to the west of the nucleus.

\subsection{The Nuclear Zone}\label{sec:nr}

We performed  X-ray spectral analysis of the diffuse nuclear emission
 defined as the light gray zone of Figure~\ref{f:banana}
 from the \nhkt\ correlation.
This zone is roughly 15\arcsec\ (650~pc) in radius.
We first fit the same absorbed thermal model  using the same energy range (0.5$-$2.0~keV) as before to validate
 the results in our thermodynamic maps.  
We obtained   \nh\ =5.8$^{+0.4}_{-0.3}\times$10$^{21}$~cm$^{-2}$,
 and $kT_e$=0.22$^{+0.01}_{-0.01}$~keV, which is consistent with the results from the
  maps.
However, with many more counts in the source
spectrum, this simple absorbed thermal model with a solar abundance is not
well fitted to the data ($\chi^2_\nu\sim 4$).
 
In fact, there are enough counts to extend the model fit to
 higher energies, 0.5$-$6.0~keV, resulting
 in 2205 $\pm$48 net counts after background subtraction.
The same local background, taken from outside the $D_{25}$ isophote,
 that was used for creating the thermodynamic map was used here.
The spectrum was grouped to contain at least 40 counts in each spectral bin,
 and the $\chi^2$ statistic was used for determining the goodness of fit.

Including the higher energy photons resulted in raising the temperature ($kT_e$=0.5) and lowering
 the column density (2$\times$10$^{21}$~cm$^{-2}$) of the
 best-fitting single-temperature model but the fit is unacceptable ($\chi^2_\nu\sim 6$).
Next, we allowed the elemental abundances in the {\tt mekal} model component
 to vary keeping their ratios fixed to the solar ratios given by \citet{anders89}.
This dramatically improved the fit to $\chi^2_\nu~\sim$1.9 with one fewer degree of freedom
and resulted in raising the best-fitting temperature to $kT_e$=0.58~keV and
  lowering \nh\ to 3$\times$10$^{20}$~cm$^{-2}$.
  
The resulting abundances were subsolar, $Z=0.05~Z_{\odot}$,
 which is inconsistent with the \xmm/RGS result \citep{perez10} and is
problematic if the hot gas originates from $\alpha$-element-rich
 core-collapse supernovae as expected.
On the other hand, the widely-applied
Weaver model \citep{castor75,weaver77} of hot bubble formation 
 predicts the majority of
 hot gas is mass-loaded from the surrounding cold
  clouds rather than coming from the metal-rich shocked supernovae gas.
It is also known that a simplistic model sometimes gives very low 
 abundances when applied to the relatively low CCD resolution
  imaging spectra \citep{weaver00}.  
Hence, we do not argue strongly that the resultant abundance estimate is realistic and
 decided to fix abundances at the solar value, consistent with \citet{perez10},
 in the remainder of our analysis.

The fitting residuals indicate that the single-temperature  model  best fits the data only between 0.7 and 1.2~keV.
Therefore, we added a second model component, either  a power-law  component
 or a second thermal model, to account for the
 softer and harder emission.
 
Adding a power-law model resulted in a significantly improved  fit statistic ($\chi^2_\nu \sim$1).
The power-law index is $\Gamma$=2.65.
 This value is similar to that found in the nuclear starburst regions of NGC~1365 \citep{wang09}
 and NGC~5135 \citep{levenson04}.
 These authors interpreted this hard emission as unresolved
  point sources; however, the power-law slope
  is much steeper than the slope 
 of the detected point sources in \ngc\ ($\Gamma\sim$1.95; see the Appendix).
 Also, fitting only the data in the range 2 and 6~keV gives an even steeper
 index of $\sim$4.
We argue that the harder X-ray emission is unlikely   
 to come from unresolved point sources and is instead a hotter thermal component.
 
\begin{figure}
\begin{center}
\includegraphics[angle=-90, width=0.45\textwidth]{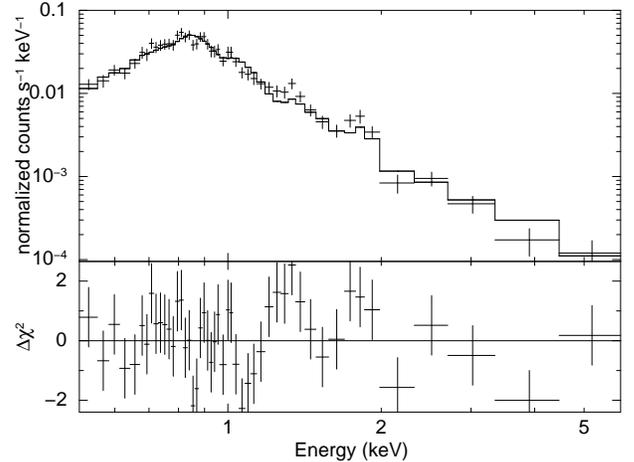} 
\caption{Spectrum of the nuclear region defined by \nhkt\
 correlation of Figure~\ref{f:banana}.  The
  spectrum was grouped to achieved at least 40 counts in each bin.  
 A background spectrum, extracted from a region to the northeast of the
 galaxy and outside of the $D_{25}$ isophote, has been subtracted.
An absorbed three temperature component {\tt mekal}
  model with solar metal abundances is shown as the solid curve.
The lower panel shows the contributions to the $\chi^2$ fit statistic.
  \label{f:spec_n}}
\end{center}
\end{figure}

Adding a second thermal component instead of a power law results in a $\chi^2_\nu \sim$2, but
not yet an acceptable fit. 
The two temperature model results in a second temperature
 of $\sim$3~keV, which may be interpreted as shocked gas directly  associated with
a ``wind fluid'' as observed in powerful starbursts like M82  \citep{strickland07,strickland09}.
This is a more realistic interpretation than the unresolved point source (power law) 
 interpretation of the hard emission for the nuclear region of \ngc.
To obtain a better fit statistic, however, a third thermal component is required.
The resultant fitting parameters are tabulated in Table~\ref{t:diffuseFits}
where the errors are quoted  at the 1$\sigma$ (68\% confidence level).
Figure~\ref{f:spec_n} shows the spectrum of the nuclear region for this 3-temperature model
 illustrating the complex line structures due to O and Fe emission between
 0.6$-$1.2~keV, Mg at 1.3~keV and Si at 1.8~keV as well as
 a hard continuum above 2~keV. Note the fit residuals indicate some of these
 line features are still poorly reproduced by this model with
 abundances fixed to their solar values.
The two lower temperatures for this model are $\sim$0.2 and 0.6~keV,
 which are typical values for star-forming regions and
the halos of starburst galaxies
\citep{strickland04, yukita10}.
This model results in a column density
 \nh=1.4$\times$10$^{21}$~cm$^{-2}$.

We also derived other physical properties of the hot gas from the spectral fits;
 namely, the electron
 density $n_e$ and pressure $P$, derived in the same manner as
 in \S~\ref{sec:tmaps}. Again, we note that the assumed geometry and hence hot gas volume can
 introduce large uncertainties on these derived properties.
In addition to these two properties, we derived the hot gas
  mass, $M_{\rm X}$ =$n_e\mu m_p V$, where $m_p$ is
  the proton mass and $\mu$=1.4 is the mass per proton, the thermal
  energy $E_{\rm th}=3n_eVkT_e$, and the cooling
 time $\tau$=$E_{\rm th}/L_{\rm bol}$.
Here, $L_{\rm bol}$ is estimated from the absorption-corrected
X-ray hot gas luminosity over the energy
 range between 0.01$-$20~keV. 
The results are listed in Table~\ref{t:hotgas}.
It has been assumed that the individual temperature components are in
 pressure equilibrium in order to estimate the electron density
 and density-dependent properties.

Finally, the X-ray spectral properties of the central region of \ngc\ can be 
compared to the properties 
 previously reported by  \citet{perez10} using a deep \xmm\ observation.
\xmm\ cannot resolve the point sources. Therefore, we fitted the  \cxo\ 
spectrum of the central 15\arcsec\ radius region {\sl including} point 
sources.
The best-fit model to the \xmm\ data is of an absorbed power law plus
  thermal component, {\tt phabs(powerlaw+mekal}),
  with power law $\Gamma$=2.1, $kT_e$= 0.5~keV, and
 $L_{\rm X(0.5-8~keV)} $ = 2.3$\times$10$^{39}$~\ergl, \citep{perez10}.
 The \cxo\ spectrum
 has higher luminosity, $L_{\rm X(0.5-8~keV)} $ = 3.2$\times$10$^{39}$~\ergl\ 
and
  is characterized better with a combination of absorbed disk blackbody and 
thermal models,
  {\tt phabs(diskbb+mekal)}, with $kT_{\rm in}$ =1.3~keV and $kT_e$= 0.5~keV.
When a power-law model is used instead of a disk blackbody, the fitted power 
law index is lower than for the \xmm\ observation,
$\Gamma=$1.8, again driven by the hard X-ray spectral shape.
Both spectra suggest the absorption column is very low, at the Galactic value, 
and that
the hot gas temperature is higher than that of the map we derived in the 
previous section.
Note that the diffuse emission only accounts for about $1/3$ of the total 
luminosity in the \cxo\ data (cf. Table 1) with the remainder contributed by 
point-like sources. The largest contribution,  $L_{\rm X(0.5-8~keV)} $ = 
1.4$\times$10$^{39}$~\ergl,  is from 
CXOJ023210.1+213008, which is also fitted well with a disk
blackbody model (see the  Appendix) with $kT_{\rm in}$=1.03~keV.
Since X-ray binaries tend to dominate the luminous X-ray point-source 
population and they are often time variable, the difference in luminosity 
estimated here compared to the \xmm\ results of  \citet{perez10} may be due simply
to variability in one or more of these bright X-ray point sources.

\begin{deluxetable}{lrrrrr}
\tablecolumns{4}
\tablewidth{0pt}
\tabletypesize{\scriptsize}
\tablenum{1} \tablecaption{Fit Parameters of X-Ray Diffuse Emission}\label{t:diffuseFits}
 \tablehead{
\colhead{Parameter} &  \colhead{Nucleus} & \colhead{West} & \colhead{East}
}\startdata
Net count     & 2448$\pm$49 &1928$\pm$53 &  2512$\pm$61\\
Fitted energy range (keV) & 0.5$-$6.0 & 0.5$-$4.0 & 0.5$-$4.0 \\
$n_{\rm H}/10^{22}$ (cm$^{-2}$)   & $0.14^{+0.04}_{-0.03}$    & $0.20^{+ 0.06}_{-0.06}$ & $0.27^{+0.05}_{-0.05}$\\
$kT_{e \rm 1}$ (keV)                     & $0.18^{+0.02}_{-0.02}$    & $0.18^{+0.01}_{-0.01}$ &  $0.19^{+0.01}_{-0.01}$\\
$K_{\rm 1}/10^{-4}$                   & $0.38^{+0.23}_{-0.27}$   &$1.58^{+0.10}_{-0.07}$ & $2.57^{+1.32}_{-0.89}$\\
$kT_{e \rm 2}$ (keV)                   & $0.55^{+0.02}_{-0.02}$     & $0.55^{+0.04}_{-0.04}$ & $0.56^{+0.03}_{-0.03}$\\
$K_{\rm 2}/10^{-4}$                 & $0.36^{+0.08}_{-0.07}$    & $0.28^{+0.05}_{-0.05}$ & $0.44^{+0.07}_{-0.06}$\\
$kT_{e \rm 3}$ (keV)                     & $3.63^{+0.81}_{-0.56}$     &  & \\
$K_{\rm 3}/10^{-4}$                 & $0.30^{+0.03}_{-0.03}$    &  & \\
$\chi^2$/dof                             &  60.1/42       & 59.2/41 & 75.36/49\\
$L^{abs}_{\rm X 0.5-8.0}/10^{38}$ (erg s$^{-1}$)  & 10.8$^{+1.1}_{-3.9}$    & 7.7$^{+0.0}_{-1.4}$ &  9.8$^{+0.0}_{-1.2}$\\
$L^{int}_{\rm X 0.5-8.0}/10^{38}$ (erg s$^{-1}$)  & 17.9 & 140 & 194\
\enddata
\end{deluxetable}

\begin{deluxetable}{lrrrrr}
\tablecolumns{4}
\tablewidth{0pt}
\tablenum{2} 
\tablecaption{Properties of X-Ray Diffuse Emission}\label{t:hotgas}
 \tablehead{
\colhead{Parameter} &  \colhead{Nucleus} & \colhead{West} & \colhead{East}
}\startdata
Volume $[f]$ (10$^{63}$ cm$^{3}$) &9.6 & 56& 67\\
$n_e[f^{-1/2}]$ (cm$^{-3}$)         &  1.07, 0.36,0.06&0.08,0.03&0.10,0.03\\
$P/k[f^{-1/2}]$   (10$^5$ K cm$^{-3}$)       &46.9& 3.5& 4.3\\
$E_{\rm th}[f^{+1/2}]$  (10$^{53}$ erg)  & 93.5&  40.4  & 57.2 \\
$M_{\rm X}[f^{+1/2}]$ (10$^5$~\msun) &7.6&32.1&44.5\\
$\tau$  (Myr) & 63.2 & 6.6  &13.0 \\
\enddata
\end{deluxetable}

\begin{figure*}
\begin{center}
\includegraphics[angle=-90, width=0.45\textwidth]{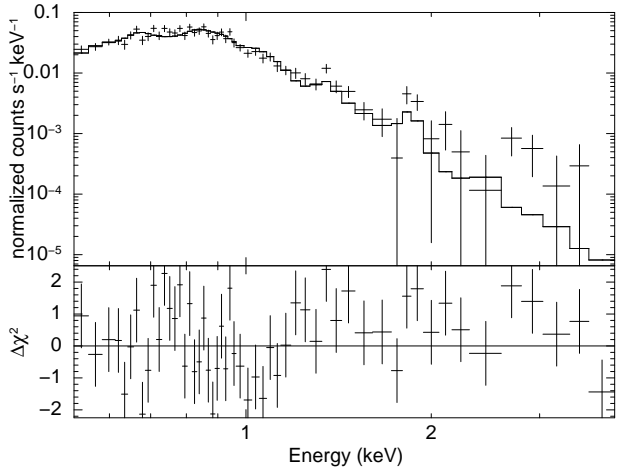} 
\hspace{0.8cm}
\includegraphics[angle=-90, width=0.45\textwidth]{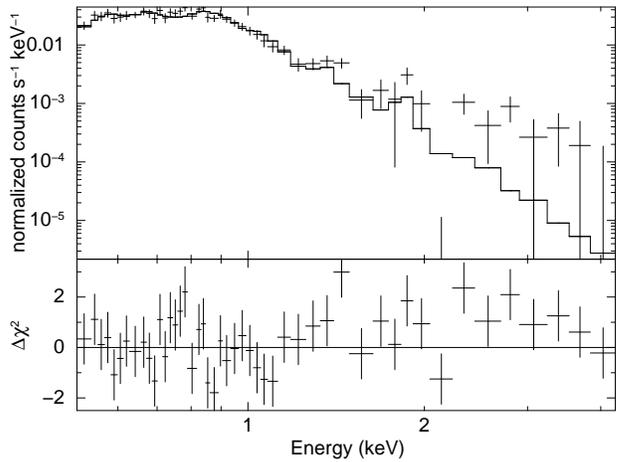}
\caption{Spectra of the east (left) and west (right) side of the low surface
 brightness diffuse emission regions fitted with the absorbed two temperature 
 models and their  the delta $\chi^2$.
  \label{f:spec_extended}} 
\end{center}
\end{figure*}

\subsection{Extended Diffuse X-ray Emission}\label{S:extendx}

Following the results of \S~\ref{sec:tmaps},
we divided the  low surface brightness diffuse emission surrounding the nuclear zone
 into two regions, namely the (roughly) east and west zones based on the
 \nhkt\ correlation map shown in Figure~\ref{f:banana}.
These zones do not show strong correlations with the underlying
 galactic structures seen at other wavelengths. However,
 the locations of dust lanes relative to the (optical) starlight from the spiral arms
 (e.g., the SDSS composite image of Figure~\ref{f:sdssx})
 suggest the eastern region lies closer along our line of sight than the western region.
Importantly, if some of the extended (non-nuclear) hot gas diffuse emission lies above (below) the
 plane of the galaxy, then we would expect the relative absorbing column
 density to be lower (higher) in the
 west (east) zone. This is indeed what is deduced from the thermodynamic maps used to define
 these two zones (Figure~\ref{f:banana}).

In order to compare to the hot gas in the nuclear starburst region,
 the same sequence of models were applied to the hot gas spectra in the east
 and west zones as was done for the nuclear region with
abundances of the thermal component fixed to solar.

We first applied the same single-temperature absorbed thermal model
as used for the thermodynamic map over the same energy range.
The spectra were grouped to have at least 40 counts in each bin, and
$\chi^2$ statistic is used.
The same background spectrum that was used above was applied.

Again, we have reproduced consistent results, specifically
\nh\ = (4.6$\pm$0.03$)\times10^{21}$~cm$^{-2}$ and $kT_e$ = 0.21$\pm$0.01~keV
 for the east region, and \nh\ = (4.3$\pm$0.03)$\times10^{21}$~cm$^{-2}$ and $kT_e$ = 0.19$\pm$0.01~keV
 for the west region.
Likewise, the fitting statistics are not acceptable ($\chi^2_\nu$ $>$2) with larger number counts in the
source spectra.
Again, there is some emission above 2~keV, hence we expanded 
the energy
 range to 0.5$-$4.0~keV for subsequent spectral analysis.

In both zones, including the higher energy range data did not change the fitting
 results significantly when fitting to a single-temperature thermal model with solar abundance.

Adding a second component (either power law or
second thermal component) improved the fit statistic
($\chi^2_\nu$ $\sim$1.4) 
in both the zones.
However, the best-fitting power law indices in both zones are
 $\Gamma\sim$ 3.7.  Such steep slopes are unphysical for unresolved
 point sources.
Therefore, we applied two-component thermal models. The fitting results are listed in Tables~\ref{t:diffuseFits} and~\ref{t:hotgas}
and the observed data, spectral model, and fit residuals are shown in
Figure~\ref{f:spec_extended}.
Adding a third thermal model component did not improve the fit
according to the F-test.

The best-fit temperatures are $\sim$0.2~keV and $\sim$0.6~keV, which are consistent with the
 two lower temperatures in the nuclear zone
 and with typical star-forming regions and
the halos of starburst galaxies
\citep{strickland04, yukita10}.
The dominant contribution is from the lower-temperature component, which is
 consistent with the single-temperature model.
Adding a second temperature component lowers the absorbing column densities in
 the two zones with \nh\ in the eastern zone slightly higher than in the
 western zone though they agree within errors.
  
\section{The Nuclear Starburst and a Search for an Active Nucleus}\label{sec:nsb}

\begin {figure*}
\begin {center}
\includegraphics [width=0.235\textwidth, angle=-90, origin=c]{f8a.eps}
\includegraphics [width=0.235\textwidth, angle=-90, origin=c]{f8b.eps}
\includegraphics [width=0.235\textwidth, angle=-90, origin=c]{f8c.eps}
\includegraphics[width=0.235\textwidth,angle=-90,origin=c]{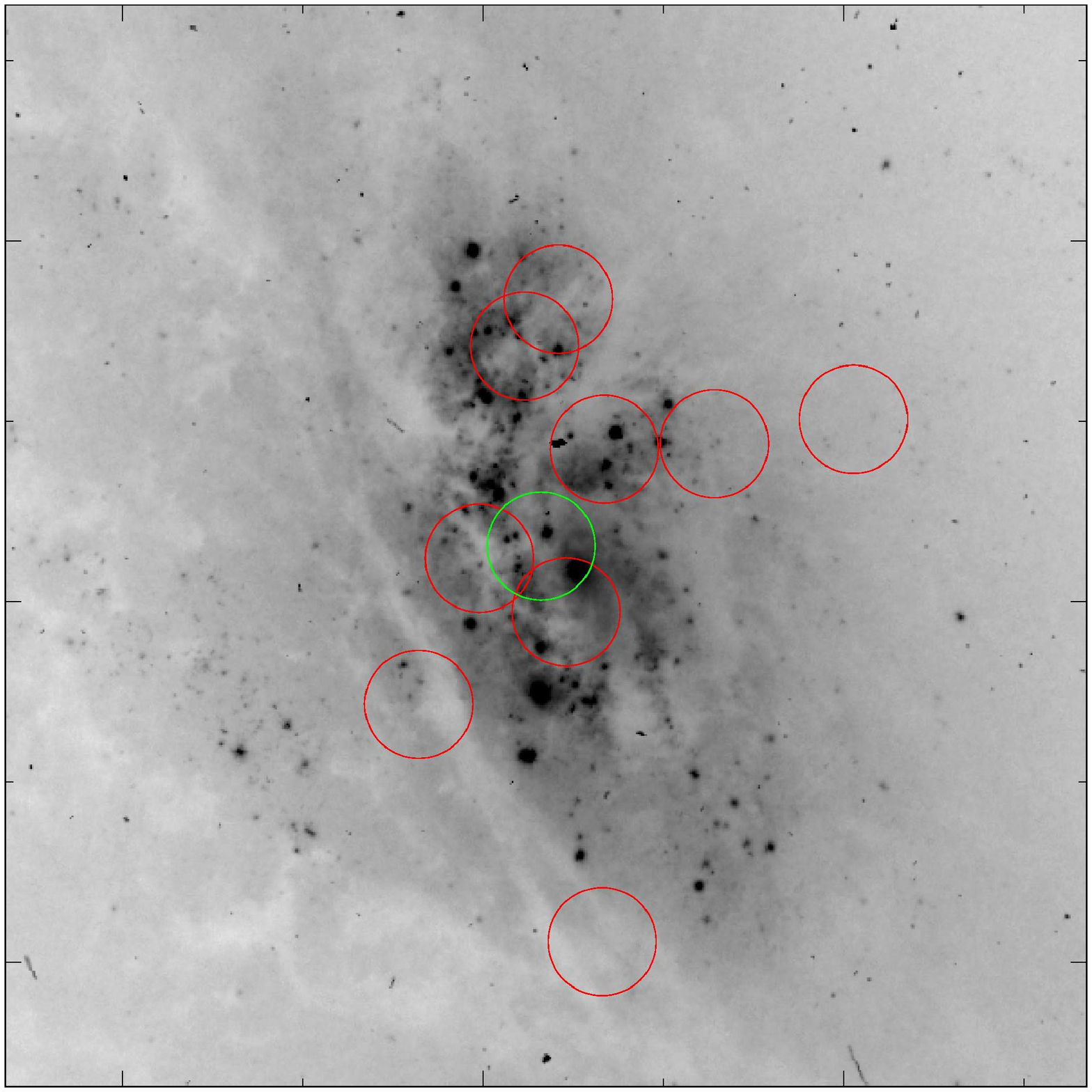}
\includegraphics[width=0.235\textwidth,angle=-90,origin=c]{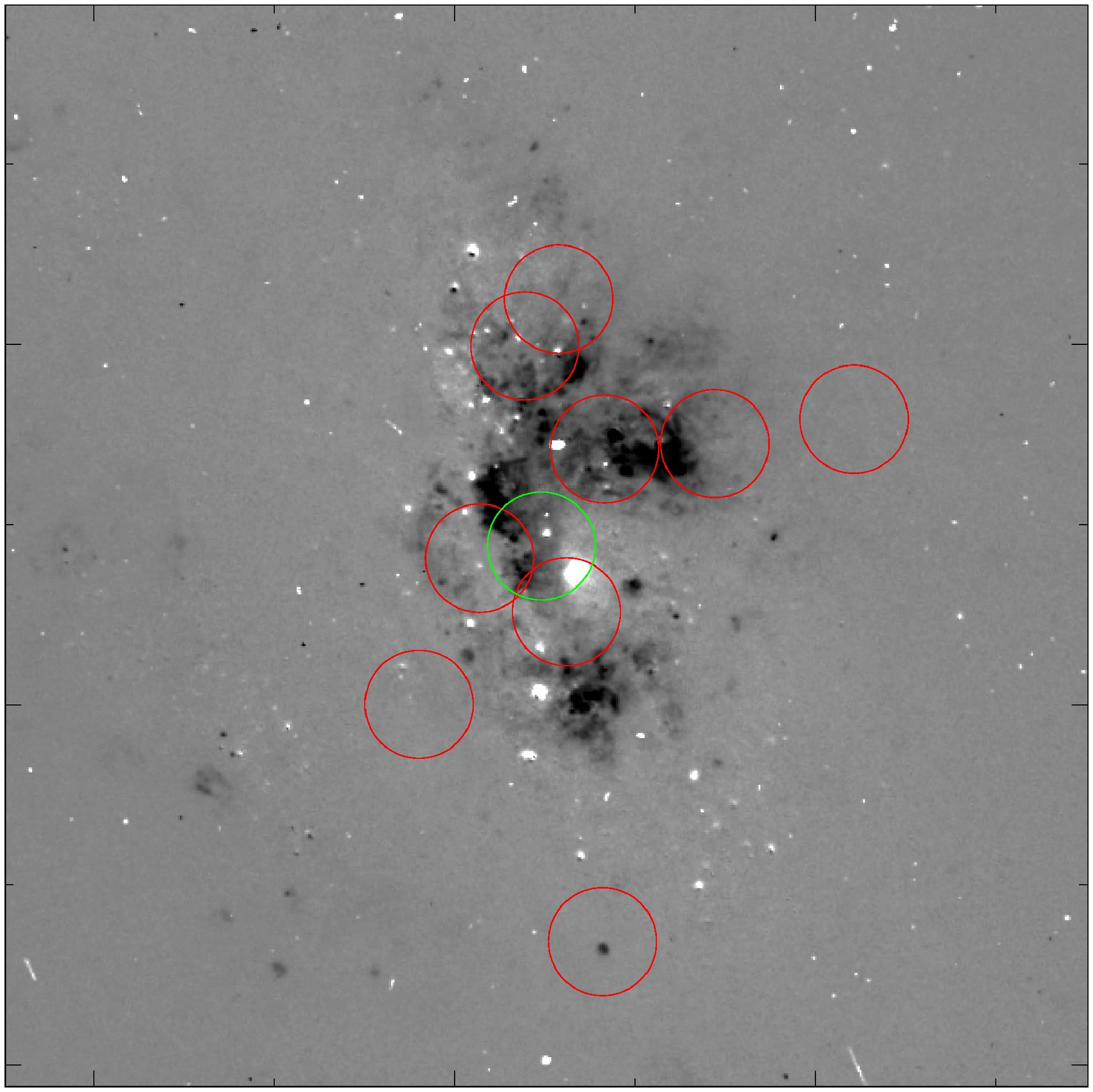}
\includegraphics[width=0.235\textwidth,angle=-90,origin=c]{f8f.eps}
\includegraphics[width=0.235\textwidth,angle=-90,origin=c]{f8g.eps}
\includegraphics[width=0.235\textwidth]{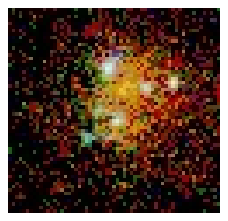}
\caption{Circumnuclear region (30\arcsec\
  $\times$ 30\arcsec) of NGC~2903.
North is up, and east is to the left.
Top: (from left to right)  the \sst\ 24~\um, \sst\ 8~\um,
{\it HST}/NICMOS 1.6~\um\, {\it HST}/ACS F814W.
Bottom: {\it HST}/ACS
continuum-subtracted \ha, \xmm\ OM UV, \cxo\ 0.5$-$8.0~keV X-ray subpixel image, and energy color-coded \cxo\
X-ray image.
The X-ray colors indicate the 0.5$-$1.0~keV (red),
 1.0$-$2.0~keV (green) and 
 2.0$-$8.0~keV (blue) energy bands.  Red circles
indicate the positions of the detected point-like X-ray sources.
The green circle depicts the position of the galaxy center as listed in
 the NASA/IPAC Extragalactic Database. The radius of a circle is 
 1\farcs5. \label{f:nuc}}
\end{center}
\end{figure*}

 Images of the nuclear region in several wavelengths are shown in Figure~\ref{f:nuc}.
Each image is 30\arcsec $\times$30\arcsec\ centered at R.A.=$9^{h}$32$^m$10\fs1,
decl.=21$^{\circ}$30\arcmin03\arcsec.
The emission is rather patchy at most wavelengths but a rough ring-like
 structure, $\sim$650~pc in diameter, has been reported in the literature
 \citep[e.g.,][]{alonso01}. Although the size of the ring is debated as one-half this size
 even after deprojection \citep{comeron10}.
The 24~\um\ image has the lowest resolution, $\sim$5\arcsec.
The 24~\um\ flux is dominated by the star-forming regions
 to the north of the center.
The 8~\um\ image shows more of a ring-like morphology and is similar to
 comparable-resolution radio images \citep[not shown; see, e.g.,][]{wynn85}.
The {\it HST}/NICOMS 1.6 \um\ and {\it HST}/ACS F814W images show
  the numerous star clusters present.
The F814W image also reveals that there are complex
 dust lanes in the east side of the nuclear starburst region.
The \ha\ image also reveals the ring-like  morphology \citep{alonso01}
 consisting of many \hii\ regions. 

The high resolution \cxo\ image of the nuclear region
 reveals that there are a number of
 point-like  sources embedded in the soft X-ray diffuse emission.
Most do not correlate spatially with features visible at comparable resolution
 at other wavelengths as expected if these sources are X-ray binaries.
Overall, the 
 X-ray morphology does not trace the  ring-like shape
 but is center-filled and circular rather than
 the elliptical shapes seen at other wavelengths that roughly match
  the inclination of the galaxy.  
The soft X-ray diffuse emission seems to be affected by the
 dust lanes, showing less emission from the east side of the center
and more absorption, appearing  green in the tri-color image
on the east edge of the starburst region where the strong dust lane
 can been seen.

As Figure~\ref{f:nuc} shows,
the mass center of \ngc\ is not well defined
 because of complex structures and the lack of an obvious active galactic signature.
NASA/IPAC Extragalactic Database lists the galactic center as R.A.=9$^{h}$32$^m$10\fs11,
decl.=21$^{\circ}$30\arcmin03\farcs0, which is obtained from the Two Micron All Sky Survey.
\citet{trachternach08} estimated the galactic center position by
fitting tilted-ring models to velocity fields derived
 from \hi\ observations (R.A.=9$^{h}$32$^m$10\fs0,
decl.=21$^{\circ}$30\arcmin02\farcs5), by fitting ellipses to the surface brightness of
 the {\it Spitzer} 3.6~\um~image (R.A.=9$^{h}$32$^m$10\fs1,
decl.=21$^{\circ}$30\arcmin04\farcs9), and by the location
 of a point source detected in the radio
 continuum (R.A.=9$^{h}$32$^m$10\fs1,
decl.=21$^{\circ}$30\arcmin04\farcs3).
We carefully registered the {\it HST}/ACS image to the 2MASS
 coordinates but found 
none of the bright star clusters or ionized regions are coincident with
 any of these four positions.
No individual X-ray point sources are detected
 at any of the four
 positions either.

We place an upper limit X-ray luminosity for any undetected source in the nuclear region
 using 1\arcsec\ source extraction radii.
The extinction toward the nuclear region is relatively moderate, reportedly ranging from
  $A_{\rm V} \sim 2-4$~mag \citep{alonso01,quillen01} for the \hii\ regions to
  $A_{\rm V} \sim$ 5.8~mag \citep{alonso01} for the average over the full 14\arcsec\ diameter
   nuclear region corresponding to hydrogen column densities of
   (4.6$-$11.2)$\times$10$^{21}$~cm$^{-2}$ based on
   the standard Galactic gas-to-dust ratio \nh/$A_{\rm V}$
    = 1.9$\times$10$^{21}$~cm$^{-2}$~mag$^{-1}$ \citep[][and references therein]{maiolino01}. (Note these values are considerably higher than our best-fit values derived
 from fits to the X-ray spectra of the nuclear region.)
A recent high resolution CO observation indicates that
 the column density can be as high as  (6$\pm$3)$\times$10$^{22}$~cm$^{-2}$ (P.-Y. Hsieh, private communication, 2011).
 Therefore, we assume a  column density of 10$^{22}$~cm$^{-2}$
to estimate a conservative upper limit X-ray luminosity from an undetected point source.
Assuming a power law spectral
 index of $\Gamma$=1.7, the 90\% confidence upper limit
 luminosity from our \cxo\ observation is  $\sim$1.5$\times$10$^{38}$~erg~s$^{-1}$.
This low value is consistent with the lack of any evidence at other wavelengths
 for an active nucleus.

\section{Discussion}\label{sec:dis}

The X-ray emission observed by \cxo\ from within the central few minutes of arc of \ngc\ 
 is composed of both hard-spectrum point-like sources
 (observed $L_{\rm X} \sim1.9\times 10^{40}$~\ergl~in the 0.5$-$8.0~keV range) and
 soft and hard diffuse extended emission (total observed  $L_{\rm X} \sim0.3\times 10^{40}$~\ergl).
The intrinsic luminosity of the detected point sources is about 25\% larger based on our
 spectral fits but the intrinsic luminosity of the soft diffuse component can be a
 factor $\sim$20 higher (Table~\ref{t:diffuseFits}) because the observed spectrum is so soft.
The observed diffuse X-ray luminosity from the central 650~pc region alone
 is $\sim$10$^{39}$~erg~s$^{-1}$ with about 30\% of this amount
 contributed by a hot component with $kT_e\sim3.6$~keV.

Much of the
 soft diffuse component is likely created by the collective effects of massive star winds and
 supernovae in star-forming regions located throughout the disk of \ngc.
We discuss below whether or not the hard diffuse component may be due
 to a concentrated hot wind fluid emanating from the central starburst region,
 the possibility of outflow from this region into the halo,
 and whether or not some portion of the soft diffuse component originates
 from this outflow.

 \subsection{Hard Diffuse X-Ray Emission}

\citet{strickland07,strickland09} have done a thorough analysis of the X-ray emission
 from the central region in the nearby starburst galaxy, M82.
We find many parallels between the X-ray properties of the central regions of \ngc\
 and those of M82, as given by Strickland \& Heckman,
 with the notable exception that \ngc\ does not so readily display
 copious galactic winds as does M82.
Part of the reason is that \ngc\ is viewed at an inclination of 60\arcdeg\ compared
 to the edge-on orientation of M82 but the wind power (or, equivalently, mass loss rate)
 is apparently much weaker in \ngc\ compared to M82.

The hard X-ray diffuse emission in M82 is confined to the nuclear starburst region
 ($r \sim 500$~pc),
 whereas the soft X-ray diffuse emission extends into the extraplanar regions above and below the
 stellar disk.
A similar distribution can be seen in NGC~2903.
The hard X-ray diffuse emission is only seen in the central \LA 15\arcsec\ (650~pc) radius
 nuclear region while
  the soft X-ray emission extends $\sim$5~kpc or more mostly to the north and west of the nucleus.
  \citet{strickland07,strickland09} also argue that the hard  continuum emission is of thermal origin;
however, as with \ngc, the hard continuum from M82 is formally better fitted with
a power law model, so that non-thermal processes such as
 inverse Compton scattering may contribute to  the hard X-ray emission.
There are too few net X-ray source counts detected from \ngc\ to
 quantify any underlying non-thermal emission component or to use emission lines
 from highly-ionized metals to better constrain the thermal component of the hard emission
 as \citet{strickland07} were able to do for M82.

The hard continuum emission temperature from M82 measured by
 \citet{strickland07} is $\sim$3.6~keV.
This is the same hard-component temperature we found for the nuclear starburst region in
 \ngc.
However, the central hot gas pressure deduced from our analysis of the X-ray spectrum
 is 5$\times$10$^6$~K~cm$^{-3}$ whereas \citet{strickland09} estimate pressures closer to
 $10^7$ to 3$\times$10$^7$~K~cm$^{-3}$ from models that reproduce the observed hard X-ray
 luminosity from M82.

Following the notation of \citet{strickland09}, we define the
 thermalization efficiency, $\epsilon$, as the fraction of mechanical energy from
 star formation that is imparted to (heats) the surrounding gas
 (the remainder is lost radiatively)
 and the mass-loading factor, $\beta$, as the ratio of the total mass of heated gas
 to the mass directly ejected by supernovae and stellar winds.
The central temperature in a starburst is then \citep[see the review by][]{veilleux05}
\begin{equation}
T_c = 0.4 \frac{\mu m_{\rm H} \epsilon L_W}{k \beta \dot{M}_W},
\end{equation}
where $\mu=1.4$ is the mean molecular weight, $m_{\rm H}$ the proton mass,
 $L_W$ the rate of energy input, $\dot{M}_W$ the rate of mass ejected from stars and supernovae,
 and $k$ is Boltzmann's constant.
Furthermore, by equating the rate of energy deposited in the surroundings,
 $\epsilon L_W$, to the asymptotic kinetic energy loss rate,
 $(1/2)\beta \dot{M}_W v_{\infty}^2$, it follows that the starburst drives a wind
 with terminal velocity
\begin{equation}
v_{\infty}^2 = \frac{2 \epsilon L_W}{ \beta \dot{M}_W } = \frac{5 k T_c}{\mu m_{\rm H}}.
\end{equation}
Since both $L_W$ and $\dot{M}_W$ scale linearly with the SFR,
 $T_c$ and $v_{\infty}$ are independent of SFR but do depend on the environment through
 $\epsilon$ and $\beta$.
Clearly, both quantities decrease as more
 cold gas mass is loaded into the wind fluid and as radiative losses increase.
Values for $L_W$ and $\dot{M}_W$ as a function of time can be taken directly from the
 Starburst99 \citep{leitherer99} stellar population synthesis models.
Using our assumption of solar metallicity
 and a Salpeter initial mass function \citep{salpeter55}
 over the stellar mass range 1$-$100~\msun, and assuming
 a continuous rate of star formation beginning at some time $t_o$,
 $L_W/\dot{M}_W$ is approximately constant after an initial transient phase lasting
 $<$20~Myr so that $T_c \sim 2.5 \times 10^8 (\epsilon/\beta)$~K.
Matching this to the observed temperature of the hot component in the
 nuclear starburst region of \ngc,
 $(\epsilon/\beta) \sim 0.15$
 so that $v_{\infty} \sim 3000 (\epsilon/\beta)^{1/2} \sim 1100$~km~s$^{-1}$.
For reasonable values of $\epsilon$ on the range 0.5--1.0
 \citep[][and references therein]{veilleux05} 
 appropriate to the moderate-density central region of \ngc,
 the mass-loading factor is relatively high, $3.3< \beta < 6.7$,
 compared to typical values derived by \citet{strickland09} from
 numerical models for M82 ($\beta \sim 2.5$).

The hard X-ray luminosity scales as
 ($\beta \dot{M}_W)^3/\epsilon R_{\star} L_W$
 \citep[see][for a derivation]{strickland09}
 where $R_{\star}$ is the radius within which the starburst energy and mass are injected. 
Taking the mean value of $\beta = 5$ for \ngc\ and $\beta = 2.5$ for M82 \citep{strickland09},
 assuming $\epsilon$ is the same for both galaxies (they likely differ by less than a factor of 2),
 and noting that $\dot{M}_W^3/ L_W \propto \rm{SFR}^2$,
 we can solve for $R_{\star}$ for \ngc\
 given the central SFRs and hard X-ray luminosities for the two galaxies
 and $R_{\star} = 300$~pc for M82 \citep{strickland09}.
This gives $R_{\star} \sim 780$~pc for the central starburst in \ngc\
 which is close to our estimated size of the star-forming region of 650~pc.
 Therefore, we conclude that the detected hard X-ray
  diffuse emission from \ngc\ is similar to that from M82
  suggesting a possible outflow originates from the central starburst
  provided the current star formation activity has persisted for $\sim$20~Myr or longer.
This implies the hot wind fluid 
present in the nuclear region of \ngc\ dominates the energetics of the flow
as is implicit in the original \citet{chevalier85} model on which the
work of Strickland \& Heckman (2009) is based.  
It is natural to ask whether or not the hot gas carries enough energy to escape
  from  the galaxy potential.
The escape velocity, $v_{e}$, is roughly 2.6$-$3.3 times the local
 circular velocity \citep[e.g.,][]{veilleux05}.
The circular velocity of the inner 7~kpc ($\sim$6\arcmin) in NGC~2903 reaches 230~km~s$^{-1}$
\citep{blok08}, yielding $v_e$ = 600$-$760~km~s$^{-1}$.
Therefore, the high-temperature component of the wind fluid is likely to flow outward and
 eventually to escape from the galaxy disk.

\begin{deluxetable}{rrrrrr}
\tablecolumns{6}
\tablewidth{0pt}
\tabletypesize{\scriptsize}
\tablenum{3} 
\tablecaption{Properties of X-Ray Point Sources}\label{t:ptsrc}.
\tablehead{
 \colhead{} &
 \colhead{R.A.} &
 \colhead{Decl.} &
 \colhead{Counts} &
 \colhead{S/N} &
 \colhead{$ L^a_{\rm X}/10^{38}$} \\
 \colhead{} &
  \colhead{(J2000)} &
  \colhead{(J2000)} &
  \colhead{} &
  \colhead{} &
  \colhead{(erg s$^{-1}$)}
}
\startdata
 1 & 09 31 54.5 &  21 27 44.6 &     22.5 &      3.8 &   0.25\\ 
 2 & 09 31 55.2 &  21 28 03.3 &     39.3 &      5.4 &   0.44 \\ 
 3 & 09 31 57.7 &  21 30 16.5 &     11.1 &      2.5 &   0.08 \\ 
 4 & 09 31 58.4 &  21 25 38.0 &      8.8 &      2.7 &   0.10 \\ 
 5 & 09 31 59.0 &  21 24 25.3 &     51.7 &      6.2 &   0.58 \\ 
 6 & 09 31 59.9 &  21 27 26.6 &     51.8 &      6.1 &   0.58 \\ 
 7 & 09 32 00.6 &  21 23 38.5 &      6.6 &      2.5 &   0.07 \\ 
 8 & 09 32 01.1 &  21 31 59.2 &      8.7 &      2.4 &   0.06 \\ 
 9 & 09 32 01.1 &  21 32 33.7 &     35.0 &      5.0 &   0.39 \\ 
 10 & 09 32 01.1 &  21 26 34.4 &     12.4 &      3.1 &   0.09 \\ 
 11 & 09 32 01.2 &  21 30 46.5 &      9.4 &      2.4 &   0.07 \\ 
 12 & 09 32 01.2 &  21 32 52.0 &    189.1 &     12.3 &   1.40 \\ 
 13 & 09 32 01.9 &  21 31 11.1 &    290.4 &     15.7 &   2.12 \\ 
 14 & 09 32 02.5 &  21 33 37.5 &      9.4 &      3.1 &   0.07 \\ 
 15 & 09 32 03.0 &  21 26 55.1 &     36.9 &      5.3 &   0.41 \\ 
 16 & 09 32 03.4 &  21 33 19.8 &     18.5 &      3.4 &   0.14 \\ 
 17 & 09 32 03.7 &  21 29 32.9 &     13.3 &      3.2 &   0.10 \\ 
 18 & 09 32 03.9 &  21 29 16.7 &     21.1 &      3.9 &   0.15 \\ 
 19 & 09 32 04.6 &  21 29 59.9 &     11.5 &      3.0 &   0.09 \\ 
 20 & 09 32 04.6 &  21 33 26.6 &    318.0 &     16.1 &   2.35 \\ 
 21 & 09 32 05.0 &  21 26 41.4 &     12.2 &      2.5 &   0.14 \\ 
 22 & 09 32 05.4 &  21 32 35.0 &   1043.0 &     29.5 &   7.68 \\ 
 23 & 09 32 06.0 &  21 28 12.8 &    157.1 &     10.8 &   1.13 \\ 
 24 & 09 32 06.1 &  21 28 18.3 &      7.3 &      2.5 &   0.05 \\ 
 25 & 09 32 06.2 &  21 30 58.7 &   2848.3 &     48.4 &  20.94 \\ 
 26 & 09 32 06.3 &  21 29 57.4 &     44.5 &      5.6 &   0.33 \\ 
 27 & 09 32 06.6 &  21 30 04.8 &     15.9 &      3.5 &   0.12 \\ 
 28 & 09 32 06.9 &  21 28 34.7 &      9.9 &      2.7 &   0.07 \\ 
 29 & 09 32 06.9 &  21 28 55.0 &     17.0 &      2.9 &   0.12 \\ 
 30 & 09 32 07.2 &  21 29 36.8 &     47.9 &      5.8 &   0.34 \\ 
 31 & 09 32 07.4 &  21 30 55.1 &     90.5 &      8.5 &   0.66 \\ 
 32 & 09 32 07.6 &  21 29 02.1 &     40.7 &      5.3 &   0.29 \\ 
 33 & 09 32 07.7 &  21 29 56.2 &     45.1 &      5.6 &   0.33 \\ 
 34 & 09 32 07.9 &  21 29 30.5 &     30.6 &      4.3 &   0.22 \\ 
 35 & 09 32 07.9 &  21 31 44.9 &      5.9 &      2.6 &   0.04 \\ 
 36 & 09 32 07.9 &  21 32 13.2 &     31.2 &      5.0 &   0.23 \\ 
 37 & 09 32 09.1 &  21 29 09.0 &    190.9 &     11.9 &   1.37 \\ 
 38 & 09 32 09.5 &  21 30 06.5 &    242.0 &     13.4 &   1.79 \\ 
 39 & 09 32 09.6 &  21 28 49.1 &     26.0 &      4.4 &   0.19 \\ 
 40 & 09 32 09.7 &  21 31 06.8 &   1234.9 &     32.0 &   8.95 \\ 
 41 & 09 32 09.7 &  21 28 36.9 &     28.4 &      4.7 &   0.20 \\ 
 42 & 09 32 09.7 &  21 24 24.1 &     18.8 &      3.6 &   0.21 \\ 
 43 & 09 32 09.7 &  21 29 06.3 &    152.1 &     10.2 &   1.09 \\ 
 44 & 09 32 09.8 &  21 30 05.8 &    208.6 &     10.6 &   1.54 \\ 
 45 & 09 32 09.8 &  21 32 17.2 &     47.0 &      6.3 &   0.35 \\ 
 46 & 09 32 09.9 &  21 34 32.5 &     30.7 &      4.4 &   0.23 \\ 
 47 & 09 32 10.0 &  21 30 05.7 &    140.3 &      6.1 &   1.04 \\ 
 48 & 09 32 10.0 &  21 29 52.0 &      8.7 &      2.7 &   0.06 \\ 
 49 & 09 32 10.1 &  21 30 01.2 &    275.5 &     13.1 &   2.03 \\ 
 50 & 09 32 10.1 &  21 30 09.8 &    103.0 &      5.8 &   1.15 \\ 
 51 & 09 32 10.1 &  21 30 08.5 &    782.4 &     22.4 &   5.82 \\ 
 52 & 09 32 10.2 &  21 30 02.7 &     93.7 &      5.5 &   0.69 \\ 
 53 & 09 32 10.4 &  21 29 58.6 &    108.8 &      8.5 &   0.80 \\ 
 54 & 09 32 10.5 &  21 31 12.0 &      9.6 &      2.6 &   0.07 \\ 
 55 & 09 32 10.8 &  21 31 26.4 &    176.1 &     12.0 &   1.31 \\ 
 56 & 09 32 10.9 &  21 29 37.7 &     12.1 &      2.7 &   0.09 \\ 
 57 & 09 32 11.5 &  21 27 41.7 &     11.4 &      3.1 &   0.08 \\ 
 58 & 09 32 11.5 &  21 30 26.1 &    137.9 &     10.5 &   1.02 \\ 
 59 & 09 32 11.8 &  21 28 34.9 &     67.7 &      7.0 &   0.49 \\ 
 60 & 09 32 11.8 &  21 24 57.3 &     46.1 &      5.6 &   0.52 \\
 61 & 09 32 11.9 &  21 27 11.2 &     29.5 &      4.7 &   0.21 \\ 
 62 & 09 32 12.0 &  21 30 33.9 &     21.3 &      4.3 &   0.15 \\ 
 63 & 09 32 12.2 &  21 27 49.6 &     40.0 &      5.4 &   0.29 \\ 
 64 & 09 32 12.3 &  21 29 22.9 &    627.1 &     21.6 &   4.59 \\ 
 65 & 09 32 12.4 &  21 30 49.5 &    110.4 &      9.6 &   0.82 \\ 
 66 & 09 32 12.5 &  21 30 25.9 &     13.1 &      3.1 &   0.09 \\ 
 67 & 09 32 12.6 &  21 24 39.9 &    346.9 &     15.9 &   3.91\\ 
 68 & 09 32 12.6 &  21 32 08.1 &     42.8 &      6.1 &   0.31 \\ 
 69 & 09 32 12.7 &  21 29 38.8 &      8.8 &      2.7 &   0.07 \\ 
 70 & 09 32 12.7 &  21 29 50.3 &     34.7 &      5.1 &   0.26 \\ 
 71 & 09 32 13.1 &  21 30 55.8 &     45.7 &      6.1 &   0.33 \\ 
 72 & 09 32 13.2 &  21 30 37.4 &     31.1 &      4.9 &   0.23 \\ 
 73 & 09 32 13.6 &  21 35 33.0 &     82.6 &      7.2 &   0.61 \\ 
 74 & 09 32 13.9 &  21 31 01.2 &     17.1 &      3.6 &   0.13 \\ 
 75 & 09 32 14.0 &  21 33 05.2 &     73.7 &      7.6 &   0.54 \\ 
 76 & 09 32 14.6 &  21 30 28.6 &     40.3 &      5.7 &   0.30 \\ 
 77 & 09 32 14.6 &  21 31 01.2 &     12.2 &      3.3 &   0.09 \\ 
 78 & 09 32 15.3 &  21 30 59.0 &      5.3 &      2.4 &   0.04 \\ 
 79 & 09 32 15.4 &  21 29 24.9 &     29.0 &      4.5 &   0.21 \\ 
 80 & 09 32 15.4 &  21 31 36.9 &     15.2 &      3.7 &   0.11 \\ 
 \enddata
\tablenotetext{a}{Intrinsic luminosity (0.5$-$8.0~keV) assuming a power law spectrum of index $\Gamma=1.95$, $ n_{\rm H}$=2.9$\times$
10$^{20}$~cm$^{-2}$.}
\end{deluxetable}

\begin{deluxetable}{rrrrrr}
\tablecolumns{6}
\tablewidth{0pt}
\tabletypesize{\scriptsize}
\tablenum{3} 
\tablecaption{Properties of X-Ray Point Sources}\label{t:ptsrc}.
\tablehead{
 \colhead{} &
 \colhead{R.A.} &
 \colhead{Decl.} &
 \colhead{Counts} &
 \colhead{S/N} &
 \colhead{$ L^a_{\rm X}/10^{38}$} \\
 \colhead{} &
  \colhead{(J2000)} &
  \colhead{(J2000)} &
  \colhead{} &
  \colhead{} &
  \colhead{(erg s$^{-1}$)}
}
\startdata
 81 & 09 32 16.4 &  21 31 11.7 &      8.9 &      2.4 &   0.07 \\ 
 82 & 09 32 17.3 &  21 29 46.5 &     77.8 &      7.9 &   0.57 \\ 
 83 & 09 32 17.7 &  21 30 20.4 &     40.5 &      5.6 &   0.30 \\ 
 84 & 09 32 18.9 &  21 26 36.1 &      6.5 &      2.4 &   0.05 \\ 
 85 & 09 32 19.6 &  21 31 05.3 &     86.0 &      8.3 &   0.63 \\ 
 86 & 09 32 20.0 &  21 27 48.9 &     18.1 &      3.8 &   0.13 \\ 
 87 & 09 32 21.0 &  21 30 02.2 &     34.7 &      5.5 &   0.26 \\ 
 88 & 09 32 21.6 &  21 28 20.9 &     19.3 &      4.0 &   0.14 \\ 
 89 & 09 32 22.2 &  21 32 41.8 &     21.9 &      3.7 &   0.16 \\ 
 90 & 09 32 22.9 &  21 29 24.7 &     34.6 &      5.2 &   0.25 \\ 
 91 & 09 32 23.0 &  21 33 02.5 &     25.1 &      4.1 &   0.19 \\ 
 92 & 09 32 23.6 &  21 29 13.7 &     12.5 &      2.9 &   0.09 \\ 
\enddata
\tablenotetext{a}{Intrinsic luminosity (0.5$-$8.0~keV) assuming a power law spectrum of index $\Gamma=1.95$, $ n_{\rm H}$=2.9$\times$
10$^{20}$~cm$^{-2}$.}
\end{deluxetable}

\begin{deluxetable*}{lrrrrr}
\tablecolumns{6}
\tablenum{4} 
\tablewidth{0pt}
\tabletypesize{\scriptsize}
\tablecaption{Fit Parameters of Bright X-Ray Point Sources}\label{t:ptfits}
\tablehead{
\colhead{Parameter} &  \colhead{J09325.4+213235} & \colhead{J093206.2+213058} & \colhead{J093209.7+213106}
& \colhead{J093210.1+213008} & \colhead{J093212.3+212922}
}\startdata
\cutinhead{MeKaL}
$n_{\rm H}/10^{22}$ (cm$^{-2}$)         & $  0.15^{+ 0.02}_{-0.02}$ & $  0.14^{+ 0.01}_{-0.01}$ & $  0.23^{+ 0.03}_{-0.03}$ & $  2.00^{+ 0.24}_{-0.25}$ & $  0.12^{+ 0.04}_{-0.04}$  \\
$kT$ (keV)                              & $  3.05^{+ 0.38}_{-0.28}$ & $  3.04^{+ 0.24}_{-0.24}$ & $  1.94^{+ 0.19}_{-0.17}$ & $  2.31^{+ 0.35}_{-0.34}$ & $  7.31^{+ 3.15}_{-1.87}$  \\
$Z/Z_{\odot}$                           & $0.00^{+0.10}_{-0.00}$ & $0.05^{+0.11}_{-0.05}$ & $0.03^{+0.05}_{-0.03}$ & $0.14^{+0.25}_{-0.14}$ & $0.00^{+ 0.57}_{-0.00}$  \\
$K/10^{-4}$                              & $  0.82^{+ 0.05}_{-0.07}$ & $  2.17^{+ 0.14}_{-0.12}$ & $  1.35^{+ 0.14}_{-0.14}$ & $  2.29^{+0.07}_{-0.04}$ & $  0.38^{+ 0.02}_{-0.5}$  \\
$\chi^2$/dof                            & 88.9/77 & 148/156 & 64.1/90 & 68.6/67 & 81.5/50 \\
$f_{\rm X 0.5-8.0 keV}/10^{-14}$ (erg s$^{-1}$ cm$^{-2}$)  & 6.67 & 18.2 & 7.04 & 7.44 & 4.74 \\
\cutinhead{Power law}
$n_{\rm H}/10^{22}$ (cm$^{-2}$)         & $  0.25^{+ 0.02}_{-0.02}$ & $  0.25^{+ 0.02}_{-0.02}$ & $  0.40^{+ 0.03}_{-0.02}$ & $  2.62^{+ 0.32}_{-0.30}$ & $  0.16^{+ 0.03}_{-0.04}$  \\
$\Gamma$                             & $  2.26^{+ 0.07}_{-0.07}$ & $  2.29^{+ 0.07}_{-0.06}$ & $  2.82^{+ 0.09}_{-0.09}$ & $  3.04^{+ 0.29}_{-0.22}$ & $  1.64^{+ 0.06}_{-0.08}$  \\
$K/10^{-4}$                             & $  0.27^{+ 0.03}_{-0.03}$ & $  0.75^{+ 0.05}_{-0.05}$ & $  0.51^{+ 0.06}_{-0.05}$ & $  1.68^{+ 0.87}_{-0.89}$ & $  0.09^{+ 0.01}_{-0.01}$  \\
$\chi^2$/dof                            & 101/78 & 151/157 & 73.3/91 & 70.9/68 & 85.6/51 \\
$f_{\rm X 0.5-8.0 keV}/10^{-14}$ (erg s$^{-1}$ cm$^{-2}$)  & 6.82 & 18.8 & 7.17 & 7.59 & 5.00 \\
\cutinhead{DiskBB}
$n_{\rm H}/10^{22}$ (cm$^{-2}$)         & $  0.05^{+ 0.02}_{-0.02}$ & $  0.05^{+ 0.01}_{-0.01}$ & $  0.12^{+ 0.02}_{-0.02}$ & $  1.61^{+ 0.38}_{-0.32}$ & $  0.04^{+ 0.03}_{-0.03}$  \\
$kT_{\rm in}$ (keV)                     & $  1.07^{+ 0.07}_{-0.05}$ & $  1.01^{+ 0.04}_{-0.04}$ & $  0.83^{+ 0.04}_{-0.04}$ & $  1.03^{+ 0.17}_{-0.14}$ & $  1.41^{+ 0.17}_{-0.11}$  \\
$K/10^{-4}$                             & $ 28.1^{+6.0}_{-6.0}$ & $ 91.4^{+25.1}_{-12.9}$ & $ 94.9^{+24.4}_{-18.9}$ & $ 67.5^{+37.8}_{-23.0}$ & $  6.03^{+ 2.02}_{-2.10}$  \\
$\chi^2$/dof                            & 84.5/78 & 171/157 & 68.9/91 & 68.2/68 & 73.5/51 \\
$f_{\rm X 0.5-8.0 keV}/10^{-14}$ (erg s$^{-1}$ cm$^{-2}$)  & 6.59 & 17.3 & 6.89 & 7.00 & 4.42 \\
\cutinhead{X-ray luminosity}
$L^{int}_{\rm X 0.5-8.0 keV}/10^{38}$ (erg s$^{-1}$) & 6.64 & 20.8 & 9.50 & 13.8 & 4.33 \\
\enddata
\tablecomments{Errors are 1$\sigma$ (68\% confidence interval) uncertainty. Intrinsic X-ray luminosity
is estimated using the best-fit model. }

\end{deluxetable*}

\subsection{Soft Diffuse X-Ray Emission}
Unlike M82, direct detection of any galactic wind emanating from the central regions of \ngc\
 is compromised due to the inclination of the galaxy disk to our line of sight.
There is soft diffuse X-ray emission with characteristic temperature 0.2$-$0.6~keV
 extending beyond the central starburst with a morphology that does not closely match
 the underlying structures seen at other wavelengths (\S \ref{sec:multi}).
Much of this gas probably originated from supernovae and stellar
  winds from massive young stars distributed throughout the stellar disk of \ngc.
Its relatively soft spectrum is likely due to dilution of hotter gas by
 thermal evaporation of surrounding cold ISM as in the standard model for
 the formation of interstellar bubbles \citep{castor75,weaver77}.
However, a portion of this soft X-ray emission may come from a galactic wind,
 also diluted through mass-loading and further cooled by adiabatic expansion,
 but lying above the disk of the galaxy.

    Based on the thermodynamic maps analyzed in \S \ref{sec:tmaps}, the soft
 diffuse X-ray emission can be
 divided into three distinctive zones, the nuclear, west and east zones.
We note that the hydrogen column density is lower in the west than in the east
zone and lowest in the nuclear zone.
This may suggest the geometry of the hot gas:
in the nuclear zone, the gas extends into the lower density halo.
The gas in the east zone is mostly confined to within the disk.
Because the galaxy is tilted so that the east side  is closest to us,
 any emission from a galactic wind along a line of sight through the east zone
  would be attenuated by the disk giving the higher
 \nh\ value.
Conversely, the hot gas observed along lines of sight through the west zone
 is likely either (also) from the disk or from the wind located above the disk
 resulting in a relatively lower absorption column density.  
This geometry naturally also explains the higher X-ray surface brightness to the west
 (Figure~\ref{f:xdiff}) but does not easily explain the observed hot gas morphology;
 the shape of the soft diffuse X-ray emission does not approximate the bi-conical
 outflow seen in M82, for instance.

In any case, the X-ray CCD data alone do not give accurate kinematic
  information of hot gas.  
The kinematics of an outflowing gas can be obtained from 
high-resolution spectroscopic measurements of warm
 ionized gas.  
In fact, \citet{hagele09} argue that H$\beta$ line structures and properties  measured from
 some nuclear star clusters are akin to those of \ha\ emission in NGC~1569 investigated by
  \citet{westmoquette07a, westmoquette07b} who concluded  that these features are
  indicative of an outflow. 
Mapping the warm ionized gas in the central region of NGC~2903 will constrain
 whether or not the hot gas is escaping from the galaxy.

 \subsection{Nuclear Activity}

There is no compact source detected coincident with the position of the
 mass center in \ngc.  This is consistent with the lack of any AGN 
 signature visible at other wavelengths.  
We place an upper limit of $L_{\rm X (0.5-8.0~keV)}$ = (1$-$2) $\times$10$^{38}$~\ergl\
 to any undetected point-like nuclear source.

We can infer the mass of the central object from the $M_{\rm BH}-\sigma$ relation \citep{ferrarese00}.
The observed stellar velocity dispersion in the nuclear region of NGC~2903 listed in the
 HYPERLEDA\footnote{http://leda.univ-lyon1.fr} database 
  \citep{paturel03} is 101.1$\pm$6.8~km~s$^{-1}$
  \citep{herudeau98}.  
This implies a black hole mass of 10$^6$--10$^7$~\msun\  \citep{tremaine02, graham11, graham12}.
Therefore, we take 10$^7$~\msun\  as an upper limit to the mass of the nucleus. 
If we assume that the ambient hot  gas is spherically accreting onto this central object, 
the volume-weighted Bondi accretion rate is estimated 
 as $\dot{M}_{\rm Bondi}$ = 5$\times$10$^{-8}$~\msun~yr$^{-1}$ using Equation (6) in \citet{soria06} and 
values of the gas temperature and density as listed in Table~\ref{t:hotgas}. 
We emphasize that this accretion rate only accounts for accretion of 
 the hot phase of gas in the nuclear region.
  
We define the dimensionless accretion parameter,  $\dot{m}$, as 
$\dot{m} \equiv 0.1 \dot{M}c^2/L_{\rm Edd} \equiv \dot{M}/\dot{M}_{\rm Edd}$, 
where $\dot{M}$ is the accretion rate and $\dot{M}_{\rm Edd}$ is the accretion rate that would produce the Eddington luminosity in a radiatively efficient case.  
For the NGC~2903 nucleus,   we obtain
$\dot{m}$ =  2$\times$10$^{-7}$.  
In such low radiative efficiency regime, accretion is likely to be advection dominated. 
From the self-similar solution of \citet{narayan94}, 
  the bolometric luminosity is estimated as  $L_{\rm bol} = \eta \dot{M} c^2 = (10 \dot{m})\dot{M} c^2$, where
   $\eta$ is the radiative efficiency.
This leads to the corresponding $L_{\rm bol}\sim$10$^{34}$~\ergl.
The predicted X-ray luminosity can be a factor of 10 lower than the bolometric
 luminosity for a supermassive black hole \citep{elvis94}, resulting in $L_{\rm X}\sim$10$^{33}$~\ergl.
A similar $L_{\rm X}$ value is also obtained from the $L_{\rm X}$--$\dot{m}$  relation 
  derived by  \citet{merloni03} for advection models  
(taking into account that the alternative definition 
of $\dot{m} \equiv \dot{M}c^2/L_{\rm Edd}$ is used in their paper.) 
The hot gas component alone can only provide enough accretion to produce $L_{\rm X}\sim$10$^{33}$~\ergl.

In general, nuclear black holes in nuclear starburst galaxies are much more active than in NGC~2903 
 because cold gas is also accreted onto the black hole.
 There is a tight correlation between nuclear SFR and 
 black hole accretion rate (BHAR)  found in Seyfert galaxies \citep{diamond12}.
The SFR-BHAR suggests that external cold gas triggers star formation and AGNs
  or nuclear star formation fuels subsequent AGN activity.
The SFR measured from hydrogen recombination lines in the central 
15\arcsec\ (650~pc) region in \ngc\ is 0.7~\msun~yr$^{-1}$ \citep{alonso01}.
For such an SFR, we would expect $\dot{M}$ $\sim$ 0.01--0.1~\msun~yr$^{-1}$ for 
 the Seyfert galaxies
 studied by \citep{diamond12} because the SFR corresponds to a cold gas inflow $\dot{M}$.
 However, given the upper limit to the X-ray luminosity, 
the cold gas $\dot{M}$ in NGC~2903 must be much lower, $\la 3\times$10$^{-5}$~\msun~yr$^{-1}$,
 and is inconsistent with the SFR-BHAR correlation, even considering
 systematic uncertainties among different methods to derive  SFRs and accretion rates.

 \citet{planesas97} estimated an H$_{2}$ mass of  1.8$\times$10$^{8}$~\msun, 
corresponding to a surface density of 160~\msun~pc$^{-2}$ in the nuclear region of \ngc.
Moreover, the recent high resolution CO image shows a gas structure
 swirling into the center (P.-Y. Hsieh, private communication).
 Therefore, there is enough material available in the central region
 to accrete on to a central massive object in NGC~2903.

We point out that only a small fraction ($\sim$ 10\%) of gas fueling nuclear SFR is needed
in order to trigger a nuclear activity as luminous as in Seyfert galaxies.  In NGC~2903,
the 10\% of cold gas inflow may be also consumed in the nuclear starburst, heated
by massive stars and supernovae, and/or ejected in a galactic outflow.

\acknowledgements
 
Support for this work was provided in part by the National 
 Aeronautics and Space Administration through Chandra Award  GO0-11099A issued by the {\it Chandra} X-ray Observatory Center, which is operated by 
the Smithsonian Astrophysical Observatory for and on behalf of the National Aeronautics 
Space Administration under contract NAS8-03060.
M.Y. acknowledges R. Buta and K. Wong for fruitful discussions.

\appendix
\section{X-ray Point-Source Analysis}

Source positions, number of source counts
 (corrected for the finite-aperture model PSF),
 S/N,
 and estimated source luminosities in the 0.5$-$8.0~keV range
 are tabulated in Table~\ref{t:ptsrc}.
The listed source positions are refinements from the
 initial source detection estimates;
refined positions were made by fitting circular
 Gaussian models to the spatial distribution of
 X-ray events in the vicinity of each source.
The source luminosities were estimated from the average
 count rate during the observation using the 
 Portable Interactive Multi-Mission Simulator (PIMMS)\footnote{http://heasarc.nasa.gov/docs/software/tools/pimms.html} 
 assuming an absorbed power-law spectral shape with spectral
 index $\Gamma=1.95$ and with a hydrogen column density
 equal to the Galactic column density
 along the line of sight of \ngc, $n_{\rm H}=2.9\times 10^{20}$~cm$^{-2}$,
 and solar abundances.
We note that estimating flux using individual response and effective area
files for each source gives more accurate value. 
However, PIMMS estimation
is reasonable here due to low S/N for the most of sources; 
therefore, the statistical uncertainties are larger than the systematic 
uncertainties due to  differences in the response matrices implemented in PIMMS 
than those of the actual observation.

Here, the X-ray spectra of the five highest-count sources are analyzed.
This includes one source in the nuclear region, CXOUJ093210.1+213008,
 the source mentioned in
 \S~\ref{S:pts}, and three others.
All five are coincident with sources
 detected with {\sl XMM-Newton} in a 21 April 2009 observation
 (P\'{e}rez-Ram\'{i}rez et al. 2010) although sources in 
the nuclear region are highly confused at the
 {\sl XMM-Newton} resolution
 (XMM-NGC2903 X-1 is 3.9\arcsec\ distant from CXOUJ093210.1+213008).
 
\begin{figure}
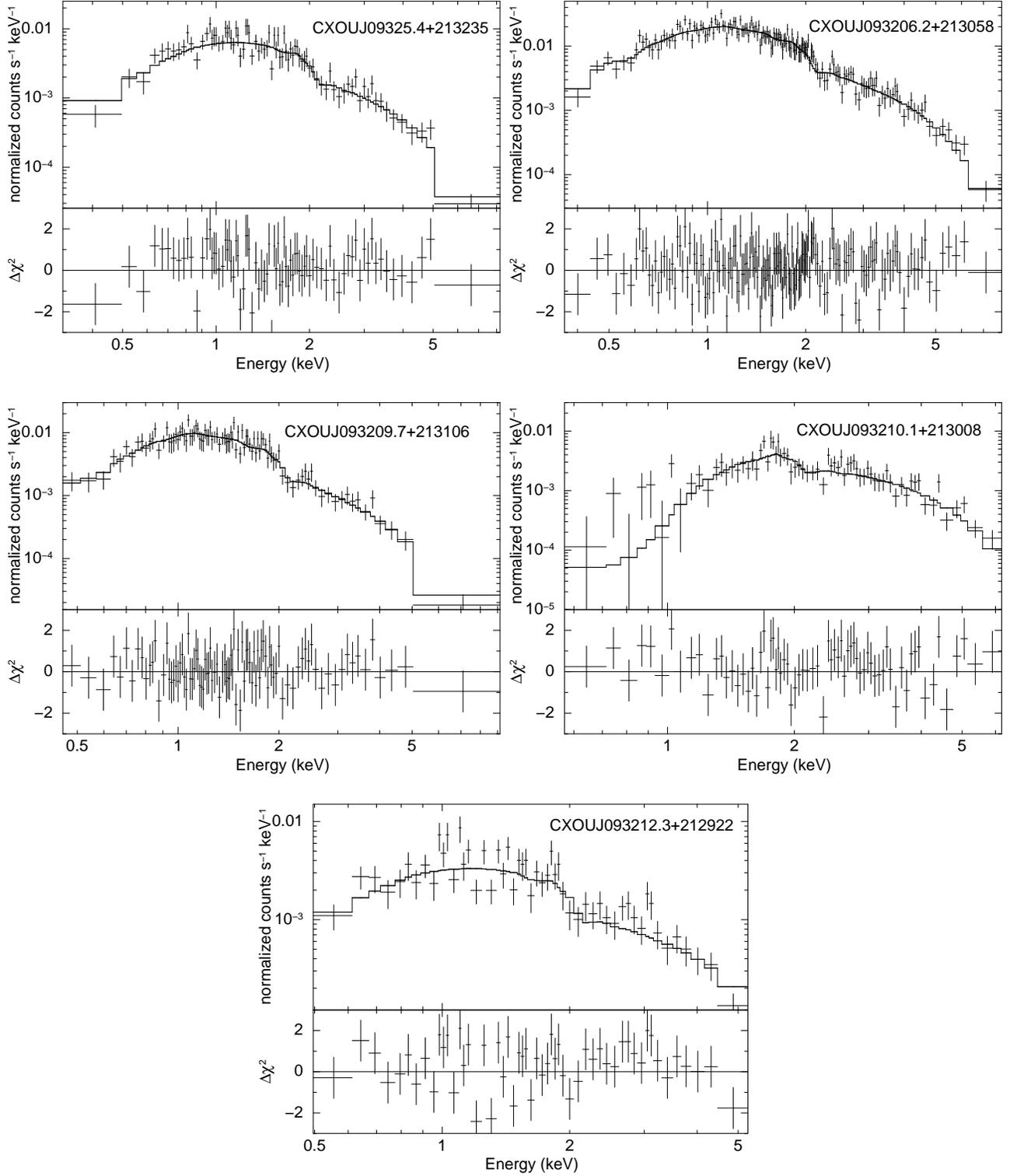

\begin{center}
\includegraphics[angle=-90,width=0.47\textwidth]{f9a.eps} \includegraphics[angle=-90,width=0.47\textwidth]{f9b.eps} \vskip 10pt
\includegraphics[angle=-90,width=0.47\textwidth]{f9c.eps} \includegraphics[angle=-90,width=0.47\textwidth]{f9d.eps} \vskip 10pt
\includegraphics[angle=-90,width=0.47\textwidth]{f9e.eps} \figcaption{Spectra, best-fitting models (solid histogram), and fit residuals (lower panel)
of the brightest sources in \ngc\ detected in the \cxo\ observation taken on March 7, 2010.
Details of all model fits including fit parameters and X-ray fluxes estimated from model fits
are listed in Table~4. 
\label{f:ptsrcSpectra}}
\end{center}
\end{figure}

Spectral analysis was performed using the {\footnotesize XSPEC} v.12.6
 spectral-fitting package using
 redistribution matrices and ancillary response files appropriate
 to the source location on the detector.
These files were generated using the {\footnotesize CIAO}
 utilities {\it mkacisrmf}, {\it mkarf}, and associated programs.
Absorbed ({\footnotesize XSPEC}'s {\tt phabs}) power law ({\tt powerlaw}),
 blackbody accretion disk ({\tt diskbb}),
 and optically-thin thermal plasma ({\tt mekal}) models
 were fitted to the observed spectra in the 0.3$-$10.0~keV range
 using the $\chi^2$ fit statistic
 (events were grouped to ensure a minimum of 10 counts per spectral energy bin).
Results of the fitting are given in Table~4 and the observed and
 model spectra and delta $\chi^2$ are displayed in
 Figure~\ref{f:ptsrcSpectra}.
The intrinsic (absorption-corrected) 0.5$-$8.0~keV luminosities of these
 sources, as estimated using the best-fitting model,
 are listed in the bottom row of Table~4.

Three sources are best fit using the blackbody-disk model.
This model is typically
 applied to accreting X-ray binary systems.
The {\tt diskbb} model results in inner disk temperatures of
 $T_{\rm in}\sim1$~keV  as expected for stellar-mass black holes \citep[e.g.,][]{remillard06}.
The most luminous source, CXOUJ093206.2+213058, and the source
 located in the bright star-forming region, CXOUJ093209.7$+$213106,
 are best fit using a thermal emission model indicative of hot gas.
Note, however, that the best-fitting abundances for these
 (and for the other three) sources are much less than the solar value
 due to the lack of observed line features.
Furthermore, the resulting plasma temperatures are much higher than is typical of
 diffuse hot gas in nearby spiral galaxies such as \ngc.
These results suggest that these two sources may also be accreting X-ray binaries.
The source in the nuclear region has by far the highest intervening
column density. This absorbing material is likely local to the nuclear region.

\end{document}